\documentclass[pdflatex,sn-mathphys-num]{sn-jnl}

\usepackage{graphicx}%
\usepackage{multirow}%
\usepackage{amsmath,amssymb,amsfonts}%
\usepackage{amsthm}%
\usepackage{mathrsfs}%
\usepackage[title]{appendix}%
\usepackage{textcomp}%
\usepackage{manyfoot}%
\usepackage{booktabs}%
\usepackage{algorithm}%
\usepackage{algorithmicx}%
\usepackage{algpseudocode}%
\usepackage{listings}%
\usepackage{pifont}
\usepackage[many]{tcolorbox}
\usepackage{fontawesome}
\usepackage{tabularx}
\usepackage{booktabs}
\usepackage[table]{xcolor}  
\usepackage{float}
\usepackage{threeparttable}
\usepackage[utf8]{inputenc}
\usepackage{makecell}

\theoremstyle{thmstyleone}%
\theoremstyle{thmstyletwo}%

\theoremstyle{thmstylethree}%

\begin{document}

\title[R$^2$ComSync: Improving Code-Comment Synchronization with In-Context Learning and Reranking]{R$^2$ComSync: Improving Code-Comment Synchronization with In-Context Learning and Reranking}


\author[1]{\fnm{Zhen} \sur{Yang}}\email{zhenyang@sdu.edu.cn}

\author[1]{\fnm{Hongyi} \sur{Lin}}\email{Hongyi.Lin@mail.sdu.edu.cn}

\author[2]{\fnm{Xiao} \sur{Yu}}\email{xiao.yu@zju.edu.cn}

\author[3]{\fnm{Jacky Wai} \sur{Keung}}\email{jacky.keung@cityu.edu.hk}
\author*[3]{\fnm{Shuo} \sur{Liu}}\email{sliu273-c@my.cityu.edu.hk}
\author[3]{\fnm{Pak Yuen Patrick} \sur{Chan}}\email{ppychan2-c@my.cityu.edu.hk}
\author[3]{\fnm{Yicheng} \sur{Sun}}\email{yicsun2-c@my.cityu.edu.hk}
\author[3]{\fnm{Fengji} \sur{Zhang}}\email{fengji.zhang@my.cityu.edu.hk}

\affil[1]{\orgdiv{School of Computer Science and Technology}, \orgname{Shandong University}, \orgaddress{\street{72 Binhai Rd}, \city{Qingdao}, \postcode{266237}, \state{Shandong}, \country{China}}}

\affil[2]{\orgdiv{The State Key Laboratory of Blockchain and Data Security}, \orgname{Zhejiang University}, \orgaddress{\street{38 Zheda Rd}, \city{Hangzhou}, \postcode{310058}, \state{Zhejiang}, \country{China}}}

\affil[3]{\orgdiv{Department of Computer Science}, \orgname{City University of Hong Kong}, \orgaddress{\street{Tat Chee Avenue}, \city{Hong Kong}, \postcode{999077}, \country{China}}}


\abstract{Code-Comment Synchronization (CCS) aims to synchronize the comments with code changes in an automated fashion, thereby significantly reducing the workload of developers during software maintenance and evolution. While previous studies have proposed various solutions that have shown success, they often exhibit limitations, such as a lack of generalization ability or the need for extensive task-specific learning resources. This motivates us to investigate the potential of Large Language Models (LLMs) in this area. 
However, a pilot analysis proves that LLMs fall short of State-Of-The-Art (SOTA) CCS approaches because (1) they lack instructive demonstrations for In-Context Learning (ICL) and (2) many correct-prone candidates are not prioritized.
To tackle the above challenges, we propose \underline{R$^2$ComSync}, an ICL-based code-\underline{Com}ment \underline{Sync}hronization approach enhanced with \underline{R}etrieval and \underline{R}e-ranking. Specifically, R$^2$ComSync carries corresponding two novelties: (1) Ensemble hybrid retrieval. It equally considers the similarity in both code-comment semantics and change patterns when retrieval, thereby creating ICL prompts with effective examples. (2) Multi-turn re-ranking strategy. We derived three significant rules through large-scale CCS sample analysis. Given the inference results of LLMs, it progressively exploits three re-ranking rules to prioritize relatively correct-prone candidates. We evaluate R$^2$ComSync using five recent LLMs on three CCS datasets covering both Java and Python programming languages, and make comparisons with five SOTA approaches. Extensive experiments demonstrate the superior performance of R$^2$ComSync against other approaches. Moreover, both quantitative and qualitative analyses provide compelling evidence that the comments synchronized by our proposal exhibit significantly higher quality.}

\keywords{Code-Comment Synchronization, Large Language Models, In-Context Learning, Multi-turn Re-ranking}



\maketitle

\section{Introduction}\label{intro}


Code comments are descriptions in natural language that record code functionalities or implementations, which are crucial to program comprehension and maintenance \cite{keyes2002software, pascarella2019classifying, sridhara2010towards}. However, with the increasing frequency of iterations of software products, developers often ignore or forget to synchronize comments when the corresponding code undergoes changes \cite{ratol2017detecting, liu2021just}. Such inconsistent or outdated comments (i.e., bad comments) hinder the program comprehension among developers and can lead to future bugs \cite{tan2007icomment,wen2019large,tan2007hotcomments,yao2024survey}. In order to eliminate or even avoid bad comments, previous studies have proposed various Code-Comment Synchronization (CCS) approaches of three categories, including heuristic-based approaches (e.g., HEBCUP \cite{lin2021automated}), deep learning-based approaches (e.g., CUP \cite{liu2020automating} and HatCUP \cite{zhu2022HatCUP}), and two-phase-based approaches (e.g., CBS \cite{yang2021significance} and Toper \cite{lin2022predictive}).
Despite the promising advances, each type of approach has its inherent limitations. For example, heuristic-based approaches try to introduce expert experience but result in a lack of generalizability \cite{yang2021significance, lin2022predictive}. Deep learning-based approaches necessitate a vast amount of high-quality code-comment co-change historical data for training, which are extremely costly to construct \cite{ni2022best, nashidretrieval}. Two-phase approaches combine the two kinds above and are constrained by their respective capability, making independent performance improvement difficult \cite{yang2021significance}.

In recent years, Large Language Models (LLMs) have been applied in various code intelligence scenarios, such as code generation \cite{jiang2023self, li2023towards}, code translation \cite{yang2024exploring,xue2025classeval,xue2025new}, and program repair \cite{xia2023automated, xue2024exploring}, owing to their powerful generalizability and exemption of task-specific retraining/finetuning during domain adaptation. As such, LLMs can naturally overcome the above limitations of current CCS approaches, but their adaptability in the CCS task has not been fully explored.
Nonetheless, we conducted a pilot analysis in Section \ref{Motivation} and found that LLMs fall short of SOTA CCS approaches across diverse Programming Languages (PLs), even with several random examples for In-Context Learning (ICL). Through a careful investigation, we summarize two primary challenges that are imperative for LLMs to address during the adaptation process in the CCS field.


\textbf{(1) Lacking instructive demonstrations for ICL-based code-comment synchronization. }

The performance of ICL is heavily dependent on the quality of prompts, particularly those demonstrations (i.e., examples) that are crucial for guiding the inference process of LLMs \cite{liu2023pre,yan2025protecting}. Our pilot analysis (shown in Section \ref{Motivation}) found that although randomly selecting examples can yield better results than zero-shot learning, it still falls short of the performance of SOTA approaches. Thus, more effective examples that can provide precise instructions are needed. 
Previous studies in fields such as code completion \cite{lu2022reacc,zhang2023repocoder}, program repair \cite{nashidretrieval}, and assertion generation \cite{nashidretrieval} have achieved significant outcomes by retrieving semantically similar examples to compose the effective prompts. However, in the context of CCS, we argue that apart from the code-comment semantic similarity, similar code-comment change patterns should also be considered when retrieval, thereby providing specific and precise editing guidance on old comments. 


\textbf{(2) Given the diverse outputs of LLMs, many correct-prone candidates are not prioritized.} 
Although the performance of LLMs in the CCS field is poor, as shown in Section \ref{Motivation}, we find that many correct candidates have been generated but not ranked first. This indicates that we should design an effective re-ranking strategy to prioritize correct-prone ones, thereby further boosting the CCS performance of LLMs. 

To overcome the challenges above, we propose \underline{R$^2$ComSync} in this paper, an ICL-based code-\underline{Com}ment \underline{Sync}hronization approach enhanced with \underline{R}etrieval and \underline{R}e-ranking.
(1) Towards curating instructive demonstrations, R$^2$ComSync first employs our proposed Ensemble Hybrid Retrieval (EHR) method to retrieve CCS demonstrations from both aspects of code-comment semantics and change patterns.
Specifically, EHR adopts CodeBERT \cite{feng-etal-2020-codebert} to encode old code, old comments, and new code, obtaining code-comment semantics. Meanwhile, it exploits eleven manually crafted expert features proposed by Lin et al. \cite{lin2022predictive} to represent the code-comment change patterns. After obtaining two demonstration pools derived from the above two aspects, respectively, we then collect the top-$(P/2)$ similar examples from each pool and concatenate them to construct a total of $P$ demonstrations, thereby ensuring the balance between semantic and expert retrieval. 
After curating a series of CCS demonstrations that are similar in both code-comment semantics and change patterns, R$^2$ComSync incorporates those retrieved demonstrations into a hard prompt, a template we formulate particularly for the ICL-based CCS approach. Subsequently, R$^2$ComSync invokes an LLM (e.g., Llama3\cite{dubey2024llama}) to conduct the code-comment synchronization under the guidance of the specific prompts. 
(2) To prioritize correct-prone candidates, we conducted an in-depth analysis of CCS samples (as shown in Section \ref{Motivation}), summarizing a series of re-ranking rules based on our observations of the relationship between code and comment co-changes. These rules are used to measure the likelihood of each candidate making mistakes and are categorized into three levels: weak, medium, and severe.
According to the error-proneness level (i.e., from weak to severe), we perform three consecutive re-rankings in order. In this way, candidates who violate the higher-level rule will be ranked lower, thereby prioritizing the candidates who are relatively more correct-prone. As an engineering innovation, R$^2$ComSync incorporates previously proven effective thoughts, using them to make domain-specific innovations that improve CCS performance for LLMs.


We evaluated R$^2$ComSync on three CCS datasets: Liu, Panth, and Pai's datasets. The former two are of Java PLs, while the last one was crawled from popular Python projects.  
Extensive experiments demonstrate that R$^2$ComSync improves the CCS performance of different LLMs (i.e., Llama3-8, Llama3-70B, and GPT-3.5-Turbo) under test by 188.98\%--694.08\% in terms of Accuracy, 83.65\%--337.92\% in terms of Recall@5, and 40.50\%--331.00\% in terms of ESS Ratio across different datasets. As for the comparison between SOTA baselines, R$^2$ComSync with Llama3-70B performs the best overall. With the best setting on Liu's dataset, it outperforms SOTA baselines on average by 31.64\% in terms of Accuracy, 13.16\% in terms of Recall@5, and 37.98\% in terms of ESS Ratio. As for Panth's dataset, R$^2$ComSync with Llama3-70B surpasses SOTA baselines on average by 142.10\%, 89.59\%, and 82.79\% in terms of Accuracy, Recall@5, and ESS Ratio, respectively. Regarding the CCS performance on Python projects, R$^2$ComSync with Llama3-70B still maintains its superiority on average by 166.98\%, 145.51\%, and 127.23\% in terms of each evaluation metric in order on Pai's dataset. 
Moreover, human evaluation in terms of similarity, naturalness, and sensitivity further confirms the superiority of R$^2$ComSync in practical usage. 
Besides, we further conduct a series of ablation experiments to study the effectiveness of our proposed EHR method and MR strategy, extend R$^2$ComSync to Qwen2.5-1.5B and Qwen2.5-Coder-1.5B to examine its applicability on small LLMs, and qualitatively discuss the superiority and limitations of R$^2$ComSync with case studies. 

The contributions of this paper can be summarized below:
\begin{itemize}
    \item We conducted the first systematic comparison between LLMs and SOTA approaches in the field of code-comment synchronization, uncovering the weaknesses of LLMs and proposing improvement directions.
    
    \item We proposed R$^2$ComSync, including an Ensemble Hybrid Retrieval (EHR) method and a Multi-turn Re-ranking (MR) strategy. The former harnesses the retrieval superiority from both semantic and expert aspects, while the latter is derived from large-scale sample analysis with high explanability.
    
    \item We conduct extensive experiments on R$^2$ComSync both quantitatively and qualitatively. Besides, we compare R$^2$ComSync with five state-of-the-art CCS approaches, showing its powerful capability. 
    
\end{itemize}

\section{Background and Related Work}\label{background}
In this section, we introduce the background and related work concerning code-comment synchronization, code-comment inconsistency detection, in-context learning for software engineering, and post-processing of large language models.

\subsection{Code-Comment Synchronization}
Code-Comment Synchronization (CCS) is an emerging research field, which aims to synchronize comments with code changes \cite{yang2021significance}. Formally, for each group of before- and after-change code snippets, namely $c$ and $c^{'}$, along with their corresponding before- and after-change comments, namely $x$ and $y$, CCS approaches are intended to devise a function that approximates $f$, i.e., $y=f(c,c^{'},x)$, where $c$, $c^{'}$, $x$, and $y$ are referred to as old code, new code, old comment, and new comment, respectively. In this paper, a CCS target is composed of a group of $c$, $c^{'}$, and $x$, whose $y$, namely reference comment, is unknown during the inference, while a CCS demonstration, as an example of a CCS target, consists of a group of $c$, $c^{'}$, $x$, and $y$. In literature, many studies have conducted various attempts to tackle this problem. 

Liu et al. \cite{liu2020automating,liu2021just} and Panthaplackel et al. \cite{panthaplackel2020learning}, for the first time, defined the CCS problem and proposed their own solutions concurrently. The former designed an LSTM-based neural network of the standard encoder-decoder paradigm, while the latter adopted a GRU-based component during their model construction and generated sequences of edits to modify old comments. Subsequently, Lin et al. \cite{lin2021automated} proposed a heuristic-based approach to update tokens/sub-tokens in old comments according to the code changes. Later, Zhu et al. \cite{zhu2022HatCUP} incorporated code structure changes into their neural network and achieved further improvement in model performance. Two latest works were proposed by Yang et al. \cite{yang2021significance} and Lin et al. \cite{lin2022predictive}, respectively. Both of them put forward a two-phase approach, i.e., identifying categories of samples before synchronizing their comments via heuristic-based or deep learning-based approaches.

However, the approaches above either lack generalizability, require tremendous task-specific training data, or cannot independently improve performance. In contrast, this paper puts forward an ICL-based CCS approach, which leverages the generalizability of LLMs without task-specific fine-tuning/retraining to make reasonable inferences for the CCS task. Consequently, the limitations of current approaches can be avoided.

\subsection{Code-Comment Inconsistency Detection}
Code-Comment Inconsistency Detection (CCID) is a closely related code task to CCS mentioned above, which aims to detect whether a code snippet can be accurately described by its associated comment \cite{tan2007icomment}. As such, CCID is a preceding task before synchronizing comments to be consistent with their code, which has been widely studied in the past several decades \cite{tan2007icomment,panthaplackel2021deep,wen2019large,rabbi2020detecting,xu2024code,10.1145/3691620.3695010,10.1145/3663529.3664458}.
Tan et al. \cite{tan2007icomment} first formulated the CCID task and proposed iComment with the combination of machine learning and program analysis techniques. Subsequently, Rabbi et al. \cite{rabbi2020detecting} designed a siamese recurrent network to analyze code and comments respectively, thereby computing their similarity for inconsistency detection purposes. Afterwards, to overcome characterization noise and labeling errors in CCS datasets, Xu et al. \cite{xu2024code} put forward MCCL that first removes noise from CCS datasets with confidence learning and then detects inconsistent code comments, thereby enhancing the model's learning ability. Very recently, CCID tools have been incorporated with LLM techniques. For example, Zhang et al. \cite{10.1145/3691620.3695010,10.1145/3663529.3664458} combined LLM-driven techniques and program analysis to identify inconsistencies in code comments, where the former was used for design constraints extraction while the latter was applied for verifying the implementation of the above extracted constraints. Their proposed approach, namely RustC$^4$, has been successfully examined on 12 large-scale real-world Rust projects.

\subsection{In-Context Learning for Software Engineering}
In-Context Learning (ICL) is an emerging learning paradigm that does not conduct parameter updates and performs the inference on LLMs directly according to the given demonstrations \cite{dong2022survey}. Inspired by its extensive application and promising results in the NLP field \cite{wei2022chain, ram2023context, chen2023stabilized, Min2022RethinkingTR}, researchers adapt ICL techniques to code-related tasks. For example, Li et al. \cite{li2023towards} adopted ICL with coding preliminaries to study automatic code generation. Nashid et al. \cite{nashidretrieval} explored the influence of demonstration retrieval on ICL-based assertion generation and program repair. Zhang et al. \cite{zhang2023repocoder} proposed an iterative retrieval-generation framework to enhance the ICL capability of LLMs in the field of code completion. 

Nonetheless, current studies using ICL did not consider task-specific adaptation during their demonstration retrieval and neglected the importance of post-processing for LLM's outputs. Therefore, we propose an ensemble hybrid retrieval method and multi-turn re-ranking strategy to augment the ICL ability of LLMs toward the CCS field. The former, tailoring to the CCS task, improves the previous retrieval methods, while the latter is used for optimizing model outputs. Both methods benefit LLMs in generating new comments of high quality and significantly improve their performance.

\subsection{Large Language Models with Post-Processing}
Large Language Models (LLMs) have demonstrated significant success in various natural language processing tasks. The diversity in their outputs enables the use of post-processing techniques to ensure the relevance and accuracy of the answers provided by the models.

Chen et al.\cite{chen2022codet} utilize the same pretrained language model to automatically generate test cases for code samples, reducing human effort and increasing the coverage of testing scenarios. Subsequently, CodeT executes the generated test cases on code samples and employs a dual execution protocol, which considers both the consistency of the output with the generated test cases and the consistency of the output with other code samples, selecting the most suitable solution from multiple samples generated by the pretrained model.
Zhang et al\cite{zhang2023coder}. introduce a collaborative framework inspired by real-world programming practices. First, an encoder model generates a series of candidate programs. Then, a reviewer model reranks these candidates by analyzing their alignment with the input instructions. The reviewer provides a complementary perspective to the encoder’s likelihood-based approach.
Naman Jain et al.\cite{jain2022jigsaw} enhance LLMs through post-processing steps based on program analysis and synthesis techniques. These techniques understand program syntax and semantics, leverage user feedback, and improve over time with usage.
Yan et al.\cite{yan2024consolidating} employ a constrained regression method to minimally adjust relevance scores generated by LLMs, aligning them with pairwise ranking preferences. The adjustments prioritize maintaining original relevance estimates while ensuring consistency with ranking predictions.

Our method requires no additional model training. We cleverly reorder generated samples based on three reranking rules derived from analyzed datasets, which have proven to significantly improve model performance

\section{Motivation}
\label{Motivation}

\begin{table}[htbp]\caption{Performances of Baselines and LLMs.}\label{Performances of Baselines and LLMs.}
\centering
\begin{tabularx}{\columnwidth}{cl *{4}{>{\centering\arraybackslash}X}}
\toprule
\textbf{Dataset}& \textbf{Approach} & \textbf{Accuracy\%$^\uparrow$} & \textbf{Recall@5\%$^\uparrow$}  & \textbf{ESS Ratio\%$^\uparrow$} \\
\midrule
\multirow{13}{*}{\shortstack{Liu's\\Dataset}}
    &CUP & $21.63_{\pm1.672}$ & $31.79_{\pm1.478}$  & $28.64_{\pm1.512}$  \\
    &HEBCUP & $26.68_{\pm0.226}$ & $27.67_{\pm0.226}$  & $27.72_{\pm0.025}$  \\
    &CBS & $28.53_{\pm0.270}$ & $35.76_{\pm1.338}$  & $31.61_{\pm0.418}$  \\
    &HatCUP & 24.30 & 35.20  & 31.33  \\
    &Toper & \textbf{30.10} & 33.04  & \textbf{34.59}  \\    
    \cmidrule(l{-0.05em}r{-0.05em}){2-5}
    &\multicolumn{4}{c}{\cellcolor{gray!20}\textbf{LLM with Zero-shot Learning}} \\
    \cmidrule(l{-0.05em}r{-0.05em}){2-5}
    &GPT3.5-turbo & $15.05_{\pm0.261}$ & $19.43_{\pm0.304}$  &$18.89_{\pm0.304}$ \\
    &Llama3-8B & $5.34_{\pm0.462}$ & $12.59_{\pm0.462}$  &$7.16_{\pm0.558}$ \\
    &Llama3-70B & $20.20_{\pm0.897}$ & $27.90_{\pm0.924}$  &$26.18_{\pm0.924}$\\
    \cmidrule(l{-0.05em}r{-0.05em}){2-5}
    &\multicolumn{4}{c}{\cellcolor{gray!20}\textbf{LLM with Two-Random-shots Learning}} \\
    \cmidrule(l{-0.05em}r{-0.05em}){2-5}
    &GPT3.5-turbo & $15.93_{\pm0.261}$ & $25.95_{\pm0.304}$  &$22.01_{\pm0.608}$ \\
    &Llama3-8B & $15.13_{\pm1.049}$ & $24.18_{\pm0.666}$  &$19.75_{\pm0.840}$ \\
    &Llama3-70B & $20.92_{\pm0.769}$ & $27.63_{\pm0.256}$  &$29.62_{\pm1.675}$\\

\midrule
\multirow{9}{*}{\shortstack{Panth's\\Dataset}}
    &CUP & $7.56_{\pm0.981}$ & $17.34_{\pm0.494}$ & $21.20_{\pm1.481}$  \\
    &HEBCUP & $8.90_{\pm0.109}$ & $9.24_{\pm0.000}$  & $19.54_{\pm0.323}$  \\
    &CBS & $11.82_{\pm0.657}$ & $17.66_{\pm0.420}$  & $25.63_{\pm0.832}$  \\     
    \cmidrule(l{-0.05em}r{-0.05em}){2-5}
    &\multicolumn{4}{c}{\cellcolor{gray!20}\textbf{LLM with Zero-shot Learning}} \\
    \cmidrule(l{-0.05em}r{-0.05em}){2-5}
    &GPT3.5-turbo & $1.59_{\pm0.495}$ & $3.31_{\pm0.495}$  &$3.96_{\pm0.495}$ \\
    &Llama3-8B & $1.19_{\pm0.374}$ & $2.78_{\pm0.857}$  &$6.35_{\pm0.561}$ \\
    &Llama3-70B & $6.75_{\pm0.857}$ & $11.51_{\pm0.561}$  &$13.49_{\pm0.000}$\\
    \cmidrule(l{-0.05em}r{-0.05em}){2-5}
    &\multicolumn{4}{c}{\cellcolor{gray!20}\textbf{LLM with Two-Random-shots Learning}} \\
    \cmidrule(l{-0.05em}r{-0.05em}){2-5}
    &GPT3.5-turbo & $9.66_{\pm0.495}$ & $13.62_{\pm0.495}$  &$15.61_{\pm0.495}$ \\
    &Llama3-8B & $5.16_{\pm0.324}$ & $10.45_{\pm0.748}$  &$12.83_{\pm1.042}$ \\
    &Llama3-70B & \cellcolor[HTML]{E1FFFF} \textbf{\boldmath$15.61_{\pm1.349}$} & \cellcolor[HTML]{E1FFFF} \textbf{\boldmath$21.69_{\pm1.309}$}  & \cellcolor[HTML]{E1FFFF} \textbf{\boldmath$27.38_{\pm0.648}$} \\
\midrule
\multirow{9}{*}{\shortstack{Pai's\\Dataset}}
    &CUP & $3.93_{\pm0.434}$ & $9.33_{\pm1.545}$ & $8.60_{\pm1.322}$  \\
    &HEBCUP & $3.00_{\pm0.000}$ & $3.00_{\pm0.000}$  & $6.53_{\pm0.186}$  \\
    &CBS & \textbf{\boldmath$6.14_{\pm0.506}$} & \textbf{\boldmath$10.73_{\pm1.254}$}   & $12.2_{\pm1.304}$  \\      
    \cmidrule(l{-0.05em}r{-0.05em}){2-5}
    &\multicolumn{4}{c}{\cellcolor{gray!20}\textbf{LLM with Zero-shot Learning}} \\
    \cmidrule(l{-0.05em}r{-0.05em}){2-5}
    &GPT3.5-turbo & $5.33_{\pm0.272}$ & $8.67_{\pm0.272}$  &$11.33_{\pm0.272}$ \\
    &Llama3-8B & $1.11_{\pm0.314}$ & $3.78_{\pm0.416}$  &$9.11_{\pm1.030}$ \\
    &Llama3-70B & $3.44_{\pm0.629}$ & $6.33_{\pm0.000}$  &\cellcolor[HTML]{E1FFFF}$12.67_{\pm0.272}$\\
    \cmidrule(l{-0.05em}r{-0.05em}){2-5}
    &\multicolumn{4}{c}{\cellcolor{gray!20}\textbf{LLM with Two-Random-shots Learning}} \\
    \cmidrule(l{-0.05em}r{-0.05em}){2-5}
    &GPT3.5-turbo & $5.33_{\pm0.272}$ & $9.00_{\pm0.272}$  &$10.29_{\pm0.272}$ \\
    &Llama3-8B & $3.00_{\pm0.720}$ & $6.78_{\pm0.567}$  &$10.56_{\pm0.416}$ \\
    &Llama3-70B & $5.44_{\pm0.416}$ & $8.78_{\pm0.157}$  & \cellcolor[HTML]{E1FFFF}\textbf{\boldmath$13.56_{\pm1.370}$} \\
\bottomrule
\end{tabularx}
\smallskip
\parbox{\textwidth}{\footnotesize \textit{Note:} The results are presented in the form of $mean_{\pm std}$, where $mean$ is the average result of 5 experiments, and $std$ is the corresponding standard deviation. The best-performing results are highlighted in \textbf{bold}. For LLMs outperforming all SOTA approaches in terms of certain metrics, the results are highlighted in a \colorbox[HTML]{E1FFFF}{light cyan} background.  ``$\uparrow$'' denotes that the higher, the better. The tables in the follow-up sections all follow this convention.}
\end{table}

\begin{figure}
    \centering
    \includegraphics[width=0.9\linewidth]{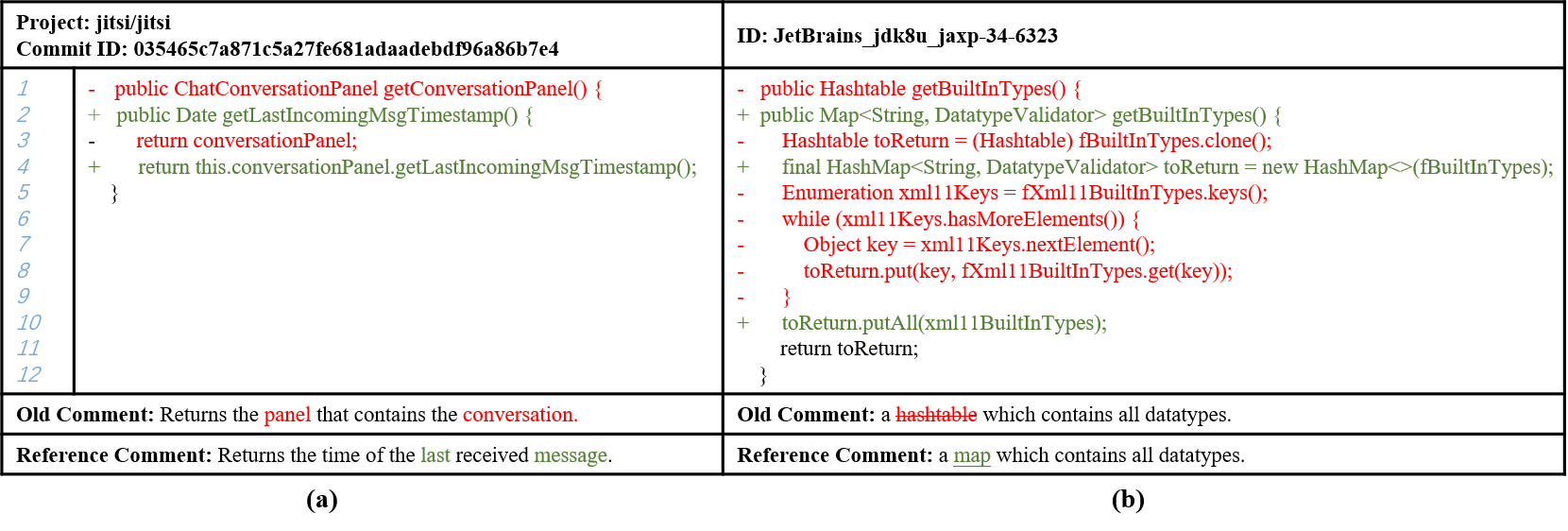}
    \caption{CCS Samples in Liu's Training Set And Panth's Traning Set}
    \label{motivation-example}
\end{figure}

Although Fan et al. \cite{fan2024exploring} have conducted an empirical study on diverse LLMs on code-comment synchronization, they did not make comparisons with the State-Of-The-Art (SOTA) methods in this field. Thus, it is unknown whether LLMs are effective tools in handling the code-comment synchronization task in practice compared with existing approaches. As such, we propose the first Research Question (RQ), namely \textbf{RQ1: How do LLMs perform against the state-of-the-art CCS approaches?}

\textbf{Experimental Design:} To address RQ1, we evaluate three LLMs, i.e., Llama3-8B, Llama3-70B, and GPT-3.5-Turbo, because they are the SOTA LLMs at the time we conduct experiments and have been widely investigated in other code comprehension studies \cite{yu2024fight,xue2024automated,sun2025source,fan2024exploring}. Besides, we measure their effectiveness according to the evaluation metrics mentioned in Section \ref{Evaluation Metrics} on all three CCS datasets introduced in Section \ref{Dataset}.
Except for Toper and HatCUP, all SOTA approaches and studied LLMs are experimented on the same testing set across all three datasets, where the former are rerun with their publicly available replication packages. Regarding Toper and HatCUP, we include them in Liu's dataset solely for comparison based on the reported experimental results in their papers. The reason is that some of their key components are not open-source. Therefore, we cannot rerun these two baselines to obtain the corresponding results on Panth and Pai’s datasets.
As for the implementation of LLMs, we follow the procedures in Section \ref{Implementaion Details}. All experiments are repeated five times. Following the above setting, we try to make a fair and reliable comparison between SOTA approaches and the studied LLMs. 

\textbf{Results:} Table \ref{Performances of Baselines and LLMs.} presents a comparative analysis between LLMs and the SOTA approaches. As for the zero-shot learning setting, only Llama3-70B obtains a slight superiority by 3.85\% against the best-performing SOTA approach (i.e., CBS) in terms of ESS ratio on Pai's dataset. For most other occasions, the results clearly indicate that LLMs struggle to compete with existing baselines. For example, the best-performing SOTA approach outperforms LLMs of this setting by 49.00\%--463.67\% in terms of Accuracy, by 26.81\%--181.02\% in terms of Recall@5, and by 32.12\%--383.10\% in terms of ESS ratio on Liu's dataset. As for Panth's dataset, the best-performing SOTA approach surpasses LLMs by 75.11\%--893.28\%, 53.43\%--535.25\%, and 89.99\%--547.22\% in terms of each metric in order. Regarding Pai's dataset, SOTA approaches still keep their superiority by 15.20\%--453.15\% in terms of Accuracy and by 23.76\%--183.86\% in terms of Recall@5.
When we turn to the two-random-shots learning setting, LLMs' performance indeed improves a lot compared with the zero-shot learning setting. But it still falls short of the performance of SOTA baselines on most occasions, as shown in Table \ref{Performances of Baselines and LLMs.}. This phenomenon illustrates that the CCS performance improvement of LLMs is limited without meaningful and relevant ICL examples. We need a retrieval module that provides effective samples for the LLMs to learn from context, thereby improving the accuracy of generated comments. 
\begin{figure}
    \centering
    \includegraphics[width=1\linewidth]{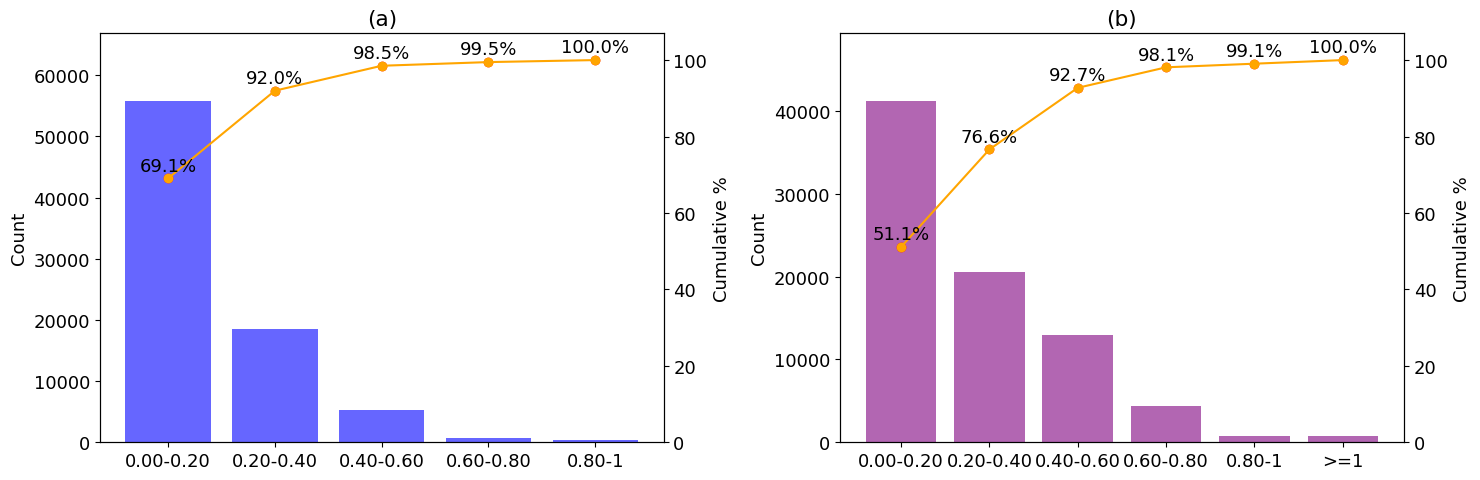}
    \caption{Sample Analysis of Liu's Training Set}
    \label{data1rule}
\end{figure}
Additionally, we have observed that many of the candidates generated by the LLMs are correct but not ranked first. This explains why the performance on Recall@5 is significantly higher than on Accuracy, both in zero-shot and two-shot learning settings. Besides, as the performance of LLMs improves, more correct answers appear within their candidate items, potentially lifting the upper bound of LLMs' CCS performance and making it possible to outperform SOTA baselines. Therefore, a re-ranking strategy is needed to prioritize the correct answers.
 
Along this line of thought, the first and second authors conducted an analysis to find out some features of correct synchronizations. 
The analysis began from the authors' prior developing experience and small-scale observation of the code-comment co-change behaviours. Subsequently, they verified these experiences and intuitive observations by large-scale analysis on Liu’s training set (a total of 80,591 samples).
First, the authors found that when a function name is changed in the code and mentioned in the corresponding old comment, 86.9\% of the training samples also change the function name in the new comment. For example, Figure \ref{motivation-example}(a) conforms to this finding exactly, where its function name in the old code is ``getConversationPanel'' while changing to ``getLastIncomingMsgTimestamp'' in the new code. Since function names normally convey the core intent of code snippets, it can be seen that the elements (e.g., ``Conversion'',``Panel'', ``Last'', and ``Msg'') involved in the function names are also changed (e.g., ``conversion'',``panel'', ``last'', and ``message'') in comments.
Next, the authors found that the proportion of sub-tokens appearing in the new comments but not in the old comments mostly accounts for 0 to 0.4 of the new comments' sub-tokens. This phenomenon occurs as frequently as 92.07\% of the total analyzed CCS samples. The detailed analysis can be seen in Figure \ref{data1rule}(a), and Figure \ref{motivation-example}(b) is a typical example that satisfies the above finding. Its new comment only has one sub-token (i.e., ``map'') that does not appear in its old comment, accounting for 0.14(1/7) of sub-tokens in its new comments.   
Finally, the authors found that the ratio of the edit distance between the new and old comments to the number of sub-tokens in the corresponding old comments mostly falls within the range of 0 to 0.6, with a sample proportion of 92.72\%. For example, the edit distance between old and new comments in Figure \ref{motivation-example}(b) is 1 (i.e., edit from ``hashtable'' to ``map''), accounting for 0.14(1/7) of the sub-tokens in its old comment. The detailed analysis can be seen in Figure \ref{data1rule}(b). 

These three findings are derived from a significant amount of the CCS samples in Liu's training set, indicating that we can leverage these findings as our re-ranking rules to further filter the candidate samples generated by the LLMs, making the synchronization more in line with user expectations. As such, the main idea of our re-ranking strategy is established, with the detailed methodology illustrated in Section \ref{sec:4.3}. It should be noted that we did not verify these rules in other datasets (e.g., Panth and Pai's) because we plan to conduct experiments (see Section 6) to examine their generalities across other datasets or programming languages, thereby demonstrating their practical value.

\newtcolorbox{boxK}{
    sharpish corners, 
    boxrule = 0pt,
    toprule = 4.5pt, 
    enhanced,
    fuzzy shadow = {0pt}{-2pt}{-0.5pt}{0.5pt}{black!35} 
}
\begin{boxK}
 \faForward \textbf{RQ1:} LLMs struggle to match state-of-the-art baselines in the code-comment synchronization task, but retrieving relevant demonstrations and re-ranking candidate new comments potentially enhance their CCS performance.
\end{boxK}

\section{Methodology}
\label{sec:4}
\begin{figure*}[htbp]
\includegraphics[width=1\textwidth]
{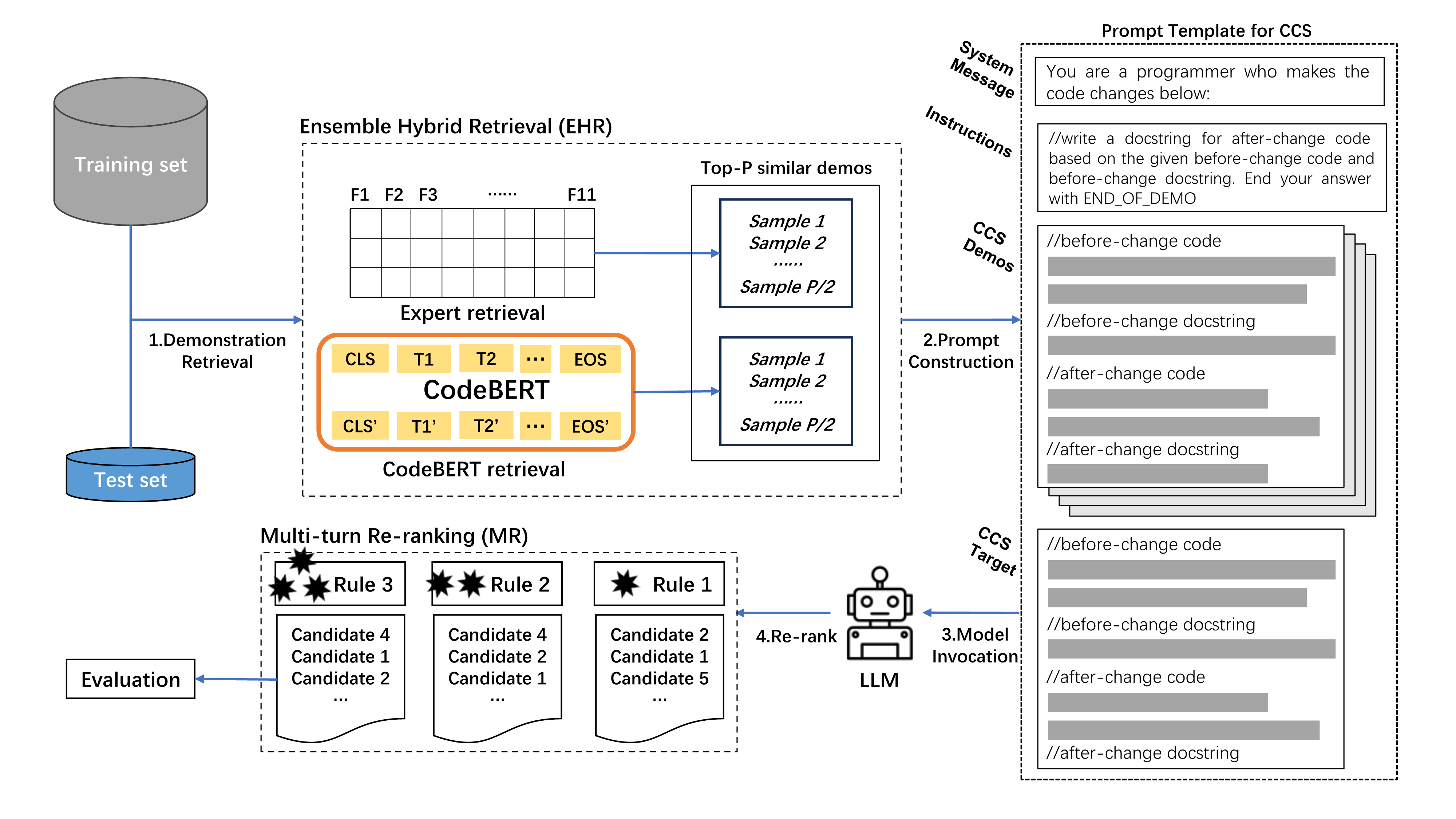}
\caption{The Overview of R$^2$ComSync.}
\label{Approach Overview}
\end{figure*}
Based on the above analysis in RQ1, we propose R$^2$ComSync with exclusively designed retrieval and re-ranking methods.
The overview of R$^2$ComSync is illustrated in Figure \ref{Approach Overview}, which totally consists of four phases: \ding{182}\ demonstration retrieval, where R$^2$ComSync employs the EHR method to retrieve similar CCS demonstrations for each CCS target from the demonstration pool (i.e., the training set). 
The EHR method takes into account both the similarities of code-comment semantics and change patterns, where the former is extracted via CodeBERT \cite{feng-etal-2020-codebert}, while the latter is derived from eleven CCS expert features manually constructed by Lin et al. \cite{lin2022predictive}; \ding{183}\ prompt construction aims to create prompts with retrieved demonstrations based on our exclusively designed template; \ding{184}\ model invocation proceeds on LLMs (e.g., Llama3), which takes prompts as inputs for ICL and generates a series of candidate new comments; and \ding{185}\ re-ranking phase exploits our proposed multi-turn re-ranking strategy to prioritize correct-prone candidates, from which we export the final outputs for evaluation. The following subsections elaborate on each phase in detail.

\begin{figure*}[htbp]
\includegraphics[width=1\textwidth]
{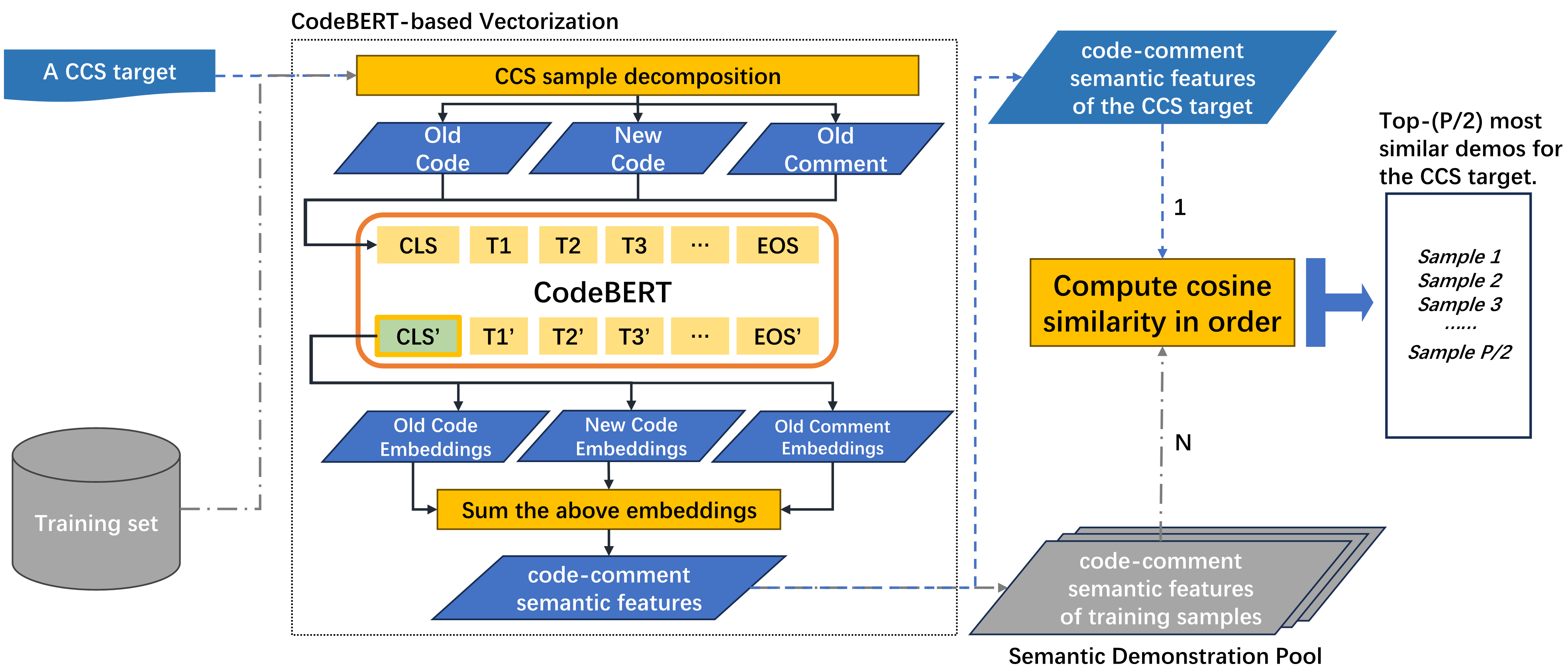}
\caption{The Illustration of CoderBERT Retrieval.}
\label{The Illustration of CoderBERT Retrieval}
\end{figure*}

\subsection{Demonstration Retrieval}
\label{sec:2}
The demonstration retrieval phase is carried out with the EHR method, retrieving a total of $P$ examples from two demonstration pools, including both code-comment semantic and change-pattern aspects. To achieve this, we design two steps, namely (1) CodeBERT retrieval and (2) Expert retrieval, taking into account both aspects accordingly.
To balance the weights of the two retrievals, we take the top-($P/2$) similar examples from each demonstration pool and concatenate these examples as the retrieval results. Detailed procedures of both CodeBERT retrieval and Expert retrieval are elaborated below.

\subsubsection{CodeBERT Retrieval.} In the community, code-related semantic feature extraction mainly relies on pre-trained LLMs for their powerful representation capability \cite{ni2022best, inala2022fault,zhang2022improving}. CodeBERT, as a prominent LLM for code representation, has demonstrated its effectiveness across a range of software engineering tasks \cite{yang2023exploitgen,nashidretrieval,lu1codexglue}, making it a good option for extracting code-comment semantic features in this paper. To be specific, we encode the old code, old comment, and new code of each sample by CodeBERT separately, owing to its encoding length limitation. Following the workaround of previous studies \cite{ni2022best, kenton2019bert}, we add the token ``[CLS]'' and the token ``[EOS]'' at the beginning and ending position of each input (i.e., the old code, old comment, and new code), and gather word embedding of token ``[CLS]'' as each of their contextual-aware representation. Afterwards, we sum the embeddings of each input and refer to the result as the code-comment semantic features. Following the above method, we vectorize all samples from the training set as a semantic demonstration pool. When retrieving demonstrations, we compute cosine similarity between the above samples and those in the test set in pairs, thereby finding the top-$(P/2)$ most semantically similar examples for each testing sample. A more explicit illustration is shown in Figure \ref{The Illustration of CoderBERT Retrieval}.

\subsubsection{Expert Retrieval.} Here, we make use of manually crafted expert features proposed by Lin et al. \cite{lin2022predictive} for representing code-comment change patterns owing to their successful application in the CCS classification task \cite{lin2022predictive}. These features (as shown in Table \ref{Expert Features that Imply Code-Comment Change Patterns}) are derived from large-scale empirical analysis based on the complexity of code changes, the extent of code changes in old comments, and the contextual information of code changes. However, it should be noted that the derivation algorithm of one of the features (i.e., TE under the context dimension, specifically the \textbf{T}ype of the \textbf{E}xpression where the changed token locates) is not open source, which makes cross-dataset extension difficult. Hence, we drop the feature TE in the follow-up experiments and retain eleven features in total, depicting code-comment change patterns from three main dimensions, including complexity, involvement, and context.  
Similarly, we perform vectorized representation based on the above features on the training sets to construct the demonstration pool. Then, we calculate the cosine similarity to find the top-$(P/2)$ similar examples from the above demonstration pool for each CCS target in the testing set. 

\begin{table}[htbp]\centering\caption{Expert Features that Imply Code-Comment Change Patterns}\label{Expert Features that Imply Code-Comment Change Patterns}
\begin{tabular}{l|l|l}
\toprule
Dim & Features & Definition \\
\midrule
\multirow{6}{*}[0pt]{Complexity} & NMS & Number of Modified Sub-tokens. \\
 & NMT & Number of Modified Tokens. \\
 & NML & Number of Modified Lines. \\
 & NMC & Number of Modified Chunks. \\
 & NNTRP & \begin{tabular}[c]{@{}l@{}} Number of Non-redundant Token  Replacement Pairs. \end{tabular}\\
 & NNSRP & \begin{tabular}[c]{@{}l@{}} Number of Non-redundant Sub-token  Replacement Pairs. \end{tabular}\\
 \midrule
\multirow{2}{*}[-5pt]{Involvement} & NTOD & \begin{tabular}[c]{@{}l@{}} Number of Tokens which Occur in the old \\ comment but Disappear after the code change. \end{tabular} \\
 & NSOD & \begin{tabular}[c]{@{}l@{}} Number of Sub-tokens which Occur in the old \\ comment but Disappear after the code change. \end{tabular} \\
 \midrule
\multirow{3}{*}[0pt]{Context} & TS\_1 & \begin{tabular}[c]{@{}l@{}} the Type of the Statement where the FIRST changed token locates. \end{tabular} \\
 & TS\_2 & \begin{tabular}[c]{@{}l@{}} the Type of the Statement where the SECOND  changed token locates. \end{tabular} \\
 & TS\_3 & \begin{tabular}[c]{@{}l@{}} the Type of the Statement where the THIRD changed token locates. \end{tabular}\\
\bottomrule
\end{tabular}
\end{table}

\medskip
\medskip
\noindent 


\subsection{Prompt Construction and Model Invocation}
\label{sec:4.2}
As shown in the right-hand side of Figure \ref{Approach Overview}, we design an exclusive prompt template for R$^2$ComSync in phase 2, which comprises a system message to designate roles, an instruction that delivers commands, and a series of placeholders (gray boxes in the prompt template) to be filled by CCS demonstrations and a CCS target. Each demonstration is separated by a delimiter ``END\_OF\_DEMO''. Towards the few-shot learning setup, given each CCS target in the testing set, we fill its corresponding retrieved demonstrations and itself in the template, while for the zero-shot learning setup, demonstrations will be ignored. 
Subsequently, in phase 3, prompts are fed into a plugged LLM (here, we use Llama3 \cite{dubey2024llama} and GPT-3.5 series) to execute the code-comment synchronization. We select the 70B parameter version of LlaMa3 as the main model for this experiment and also replicate our experiments on other LLMs, such as LlaMa3-8B and GPT-3.5-Turbo, to verify the results.

\subsection{Multi-turn Re-ranking}
\label{sec:4.3}
For each CCS target, given a series of candidate new comments generated from LLMs, we conduct the Multi-turn Re-ranking (MR) strategy to prioritize correct-prone ones. In detail, we design three heuristic rules, ranging from weak to severe, based on our pilot analysis in Section \ref{Motivation}. We execute the above rules from weak to severe for each candidate and place them in the last position if they violate the rules. Since the candidate order can be changed after each execution of one of the above rules, the worst candidate that violates the most rules will be ranked last. In contrast, the relatively correct-prone candidates can be prioritized passively. If multiple candidates violate the same rule simultaneously, they will be ranked last together with their original relative order. A MR strategy example and its illustration are shown in Figure \ref{Case Study 3} of Section \ref{Case Study}. Below, we introduce the specific mechanism of each heuristic rule.


\begin{itemize}
    \item \textbf{Rule 1 (Weak):} If the function name is changed and the changed contents can be found in the old comment, the corresponding changes should also appear in the candidate new comment as well. 
    
    \textbf{Explanation:} Typically, a function name contains the main idea of a code snippet \cite{gao2019neural}, and comment updates should be sensitive to its changes. This phenomenon accounts for 86.9\% of the CCS samples as testified in Section \ref{Motivation}. To this end, once the function name has been changed, but the corresponding contents in the old comment were not updated, the candidate new comment is likely to be incorrect. Since this rule only covers instances with function name changes and normally concerns mistakes of only several sub-tokens, we label it as a weak level. A typical example can be found in Figure \ref{Case Study 3} of Section \ref{Case Study}, which is a CCS sample in Liu's dataset. When Rule 1 is executed as the first rule, the third candidate (i.e., ``Is the counter valid for usage? '') is ranked in the last position, because the token ``valid'' in the old comment is not replaced by ``active'' as the code change occurs in their function names. Since other candidates do not violate Rule 1, the candidates' order after executing Rule 1 is: candidate 1, 2, 4, and 3.
    
    \item \textbf{Rule 2 (Medium):} 
    The number of sub-tokens in a candidate new comment but not in the corresponding old comment ($N_{diff}$) should not exceed a predefined threshold ($\sigma$) relative to the number of sub-tokens in the candidate new comment ($N_{cm^{'}_{s}}$). 
    Formally, Rule 2 is defined as below:
    \begin{equation}
     N_{diff} / N_{cm^{'}_{s}} < \sigma,
    \end{equation}
    \textbf{Explanation:} As we analyzed in Section \ref{Motivation}, comment updating normally occurs with limited ranges. For example, new comments of 92.07\% of CCS samples produce less than 40\% of new sub-tokens against old comments.
    Thus, constraining the ratio of such sub-tokens is reasonable. Since this rule only considers differences in sub-token values between old comments and candidate new comments, we label it as medium level. Figure \ref{Case Study 3} of Section \ref{Case Study} can also be a typical example to illustrate Rule 2. As can be seen, the value of $N_{diff} / N_{cm^{'}_{s}}$ for candidates 1-4 are 0.2(1/5)$<$0.35\footnote{Rule 2 ($\sigma$) is set to 0.35 for CCS samples of Liu's dataset as mentioned in Section \ref{Implementaion Details}.}, 0.14(1/7)$<$0.35, 0(0/7)$<$0.35, and 0.56(5/9)$>$0.35. Hence, candidate 4 violates Rule 2 here and should be ranked last based on the order in the previous round. As such, the candidates' order after executing Rule 2 becomes: candidate 1, 2, 3, 4.

    \item \textbf{Rule 3 (Severe):} The ratio between the comment edit distance ($ED_{cm}$) and the number of sub-tokens in the corresponding old comment ($N_{cm_{s}}$) should not exceed the predefined threshold $\epsilon$, where $ED_{cm}$ is computed by the word-level edit distance \cite{navarro2001guided} between old comments and candidate new comments. Formally, Rule 3 is defined as below:
    \begin{equation}
     ED_{cm} / N_{cm_{s}} < \epsilon,
    \end{equation}
    \textbf{Explanation:} As revealed in Section \ref{Motivation}, if the threshold of Rule 3 is 0.6, 92.72\% of CCS samples in the training set are satisfied, illustrating that we should impose restrictions on edit distance between old comments and candidates. Because the randomness during the decoding process of LLMs may potentially cause the generated sub-tokens to be out of order \cite{xu2022systematic} or irrelevant to the CCS target, leading to a high value on Rule 3. Since the computation of edit distance considers differences in both the orders and values of sub-tokens, we label this rule as a severe level. Similarly, we continue taking the example in Figure \ref{Case Study 3} for illustration. The value of $ED_{cm} / N_{cm_{s}}$ for candidates 1-4 are 0.429(3/7)$>$0.25\footnote{Rule 3 ($\epsilon$) is set to 0.25 for CCS samples of Liu's dataset as mentioned in Section \ref{Implementaion Details}.}, 0.14(1/7)$<$0.25, 0(0/7)$<$0.25, and 0.714(5/7)$>$0.25. Hence, candidates 1 and 4 both violate Rule 3 here and should be ranked last again, while keeping their original relative orders. As such, the candidates' order after executing Rule 3 becomes: candidate 2, 3, 1, 4, which is the final order (With MR Strategy) shown in Figure \ref{Case Study 3}.
\end{itemize}

\section{Experiment Preparation}
\label{sec:1}
This section presents the dataset, baselines for comparison, evaluation metrics, and implementation details of LLMs and R$^2$ComSync.
\subsection{Dataset}
\label{Dataset}
For experiments, we use two Java datasets and a Python dataset to assess code-comment synchronization, thereby examining the effectiveness of each approach comprehensively.
The first dataset is extracted from 1,496 popular Java code repositories hosted on GitHub \cite{GitHub35:online}, which has been widely used in this field \cite{liu2020automating, lin2021automated, zhu2022HatCUP, yang2021significance, lin2022predictive} and was successively polished by Liu et al. \cite{liu2020automating} and Lin et al. \cite{lin2021automated}. The dataset contains a total of 80,591 training samples, 8,827 validation samples, and 9,204 testing samples. As it was first used by Liu et al. \cite{liu2020automating} for code-comment synchronization, we refer to it as Liu's dataset in this work.
The second dataset, extracted by Panthaplackel et al. \cite{panthaplackel2020learning} from popular open-source Java projects, contains a total of 5,791 training samples, 712 validation samples, and 736 testing samples. To distinguish from Liu's dataset, we refer to it as Panth's dataset.
Furthermore, to investigate whether our method can be extended to other programming languages, we introduce a third dataset crawled from Python projects. This dataset is adapted from the CoDocBench proposed by Pai et al. \cite{11025763}, which originally contains 4,573 Python samples. 
Since the original dataset includes some samples with excessively long comments, which make existing SOTA approaches (e.g., CUP) hard to converge during training, we remove samples with comments exceeding the median length. From the remaining data, we randomly select 300 samples as the testing set and retain 3,013 samples as the training set. 
Consistent with the naming convention of the previous datasets, we refer to this dataset as Pai’s dataset. 

For each of the datasets above, all samples consist of two code-comment pairs of the old and new versions for CCS purposes. Table \ref{tab:dataset_stats} presents their detailed characteristics, where both their code and comments are tokenized into sequences and further decomposed into sub-tokens based on CamelCase and snake\_case naming conventions. The metric Length indicates the count of sub-tokens in each sequence, while the Number of Changed Sub-tokens (NCS) quantifies the modified sub-tokens between the old and new versions of code (or comments). To compromise the limited computational resources while ensuring the validity of our experiments, we sample 368 data points from the testing set of Liu’s dataset with a 95\% confidence level for the performance evaluation. In contrast, the other two datasets are kept intact for experiments. The training sets of all three datasets are utilized as retrieval databases. 

\begin{table}[ht]
\centering
\caption{Statistics of Datasets}
\label{tab:dataset_stats}
\begin{tabular}{@{}c@{\hskip 4pt}l@{\hskip 4pt}lcccccc@{}}
\toprule
\multirow{1}{*}{\textbf{Dataset}} & \multirow{2}{*}{\textbf{}} & \multirow{2}{*}{\textbf{Type}} & \multicolumn{3}{c}{\textbf{Length}} & \multicolumn{3}{c}{\textbf{NCS}} \\
\cmidrule(lr){4-6} \cmidrule(lr){7-9}
\multicolumn{1}{c}{\textbf{(PL)}} & & & Mean & Std Dev & Median & Mean & Std Dev & Median \\
\midrule
\multirow{13}{*}{\shortstack{Liu's\\Dataset\\(Java)}} 
    & \multirow{4}{*}{Train} 
        & Old Code    & 91.22 & 84.42 & 60.0 & \multirow{2}{*}{34.73} & \multirow{2}{*}{57.49} & \multirow{2}{*}{11.0} \\
        & & New Code    & 91.26 & 84.98 & 58.0 & & & \\
        & & Old Comment & 14.43 & 7.34  & 13.0 & \multirow{2}{*}{3.81} & \multirow{2}{*}{3.01} & \multirow{2}{*}{3.0} \\
        & & New Comment & 14.47 & 7.63  & 13.0 & & & \\
    \cmidrule(lr){2-9}
    & \multirow{4}{*}{Validation} 
        & Old Code    & 89.60 & 82.61 & 59.0 & \multirow{2}{*}{32.73} & \multirow{2}{*}{54.66} & \multirow{2}{*}{12.0} \\
        & & New Code    & 88.77 & 82.66 & 57.0 & & & \\
        & & Old Comment & 15.16 & 7.57  & 14.0 & \multirow{2}{*}{3.83} & \multirow{2}{*}{3.03} & \multirow{2}{*}{3.0} \\
        & & New Comment & 15.21 & 7.83  & 14.0 & & & \\
    \cmidrule(lr){2-9}
    & \multirow{4}{*}{Test} 
        & Old Code    & 97.61 & 86.92 & 64.0 & \multirow{2}{*}{36.85} & \multirow{2}{*}{63.73} & \multirow{2}{*}{12.0} \\
        & & New Code    & 97.44 & 87.94 & 63.0 & & & \\
        & & Old Comment & 15.37 & 7.52  & 14.0 & \multirow{2}{*}{3.97} & \multirow{2}{*}{3.11} & \multirow{2}{*}{3.0} \\
        & & New Comment & 15.29 & 7.77  & 14.0 & & & \\
\midrule
\multirow{13}{*}{\shortstack{Panth's\\Dataset\\(Java)}} 
    & \multirow{4}{*}{Train} 
        & Old Code    & 84.19 & 124.18 & 45.0 & \multirow{2}{*}{43.31} & \multirow{2}{*}{109.2} & \multirow{2}{*}{11.0} \\
        & & New Code    & 88.67 & 125.44 & 48.0 & & & \\
        & & Old Comment & 10.33 & 9.18  & 8.0 & \multirow{2}{*}{2.42} & \multirow{2}{*}{3.37} & \multirow{2}{*}{1.0} \\
        & & New Comment & 11.30 & 9.61  & 9.0 & & & \\
    \cmidrule(lr){2-9}
    & \multirow{4}{*}{Validation} 
        & Old Code    & 88.13 & 121.75 & 47.0 & \multirow{2}{*}{39.19} & \multirow{2}{*}{77.44} & \multirow{2}{*}{12.0} \\
        & & New Code    & 86.63 & 119.36 & 51.0 & & & \\
        & & Old Comment & 10.81 & 9.15  & 8.0 & \multirow{2}{*}{2.60} & \multirow{2}{*}{4.24} & \multirow{2}{*}{1.0} \\
        & & New Comment & 11.64 & 9.65  & 9.0 & & & \\
    \cmidrule(lr){2-9}
    & \multirow{4}{*}{Test} 
        & Old Code    & 97.73 & 158.3 & 48.0 & \multirow{2}{*}{48.4} & \multirow{2}{*}{102.10} & \multirow{2}{*}{14.0} \\
        & & New Code    & 97.16 & 147.35 & 53.0 & & & \\
        & & Old Comment & 10.76 & 8.30  & 8.0 & \multirow{2}{*}{2.48} & \multirow{2}{*}{3.38} & \multirow{2}{*}{1.0} \\
        & & New Comment & 11.40 & 8.84  & 9.0 & & & \\
\midrule
\multirow{9}{*}{\shortstack{Pai's\\Dataset\\(Python)}} 
    & \multirow{4}{*}{Train} 
        & Old Code    & 156.05 & 183.33 & 97.00 & \multirow{2}{*}{5.31} & \multirow{2}{*}{5.97} & \multirow{2}{*}{4.00} \\
        & & New Code    & 166.38 & 188.29 & 108.00 & & & \\
        & & Old Comment & 43.89 & 32.91  & 37.00 & \multirow{2}{*}{2.82} & \multirow{2}{*}{2.83} & \multirow{2}{*}{2.0} \\
        & & New Comment & 56.80 & 50.08  & 45.0 & & & \\
    \cmidrule(lr){2-9}
    & \multirow{4}{*}{Test} 
        & Old Code    & 165.70 & 184.17 & 112.0 & \multirow{2}{*}{5.72} & \multirow{2}{*}{6.64} & \multirow{2}{*}{4.0} \\
        & & New Code    & 171.88 & 192.27 & 112.0 & & & \\
        & & Old Comment & 49.58 & 34.18  & 44.50 & \multirow{2}{*}{2.59} & \multirow{2}{*}{2.61} & \multirow{2}{*}{1.0} \\
        & & New Comment & 60.52 & 48.22  & 51.0 & & & \\
\bottomrule
\end{tabular}
\smallskip
\parbox{\textwidth}{\footnotesize \textit{Note:} \textbf{Std Dev} stands for standard deviation, which reflects the degree of data dispersion.}
\end{table}

\subsection{Baselines}
\label{State-of-the-art Baselines}
We comprehensively compare the performance of R$^2$ComSync with the five state-of-the-art approaches, namely CUP \cite{liu2020automating}, HEBCUP \cite{lin2021automated}, HatCUP \cite{zhu2022HatCUP}, CBS \cite{yang2021significance}, and Toper \cite{lin2022predictive}. Their detailed information is listed below.

\textbf{(1) CUP} is an LSTM-based encoder-decoder structured neural network, constructed by layers of embedding, co-attention, modeling, and pointer generator \cite{see2017get}. By effectively modeling the connections between comments and code changes, CUP has obtained promising results in the CCS field.

\textbf{(2) HEBCUP} is the first heuristic-based CCS approach, leveraging a series of fine-designed heuristic rules to conduct token/sub-token level updates on old comments. However, it is only useful for code-indicative samples, whose changed tokens/sub-tokens in comments can also be found in corresponding code changes.   

\textbf{(3) HatCUP} takes code structure change information into account. Combining data flow dependencies and code change graphs, HatCUP introduces a structure-guided attention mechanism to analyze hybrid information. Besides, it generates edit actions for old comments, instead of complete new comments, leading to its updates being more accurate. 

\textbf{(4) CBS} is a two-phase approach in the CCS field, which first predicts the proneness of each sample, i.e., heuristic-prone or non-heuristic-prone, then allocates samples to CUP or HEBCUP for code-comment synchronization according to their classification results.

\textbf{(5) Toper} is another two-phase approach concurrently proposed with CBS. Inside Toper, an Abstract Syntax Tree (AST) path-based comment updater is proposed to deal with non-code-indicative samples, while code-indicative samples are allocated to HEBCUP to handle. The whole mechanism is also operated via the process of classifying samples first, followed by synchronizing.

\subsection{Evaluation Metrics}
\label{Evaluation Metrics}
Following the previous studies in the CCS field \cite{liu2020automating, lin2021automated, zhu2022HatCUP, yang2021significance, lin2022predictive}, we adopt Accuracy, Recall@5, and ESS Ratio to evaluate the performance of each approach.

\textbf{Accuracy} refers to the ratio of CCS targets where \textit{correct synchronization} can be generated on the first attempt among all examined cases.
Particularly, \textit{correct synchronization} denotes that the generated new comments are exactly identical to their corresponding reference comments. 

\textbf{Recall@5} is a similar metric to Accuracy, but allows CCS approaches to carry out five attempts for each case examined. If any one of its attempts reaches the \textit{correct synchronization}, it considers the approach can successfully handle this case.

\textbf{ESS Ratio} evaluates the ratio of samples whose word-level edit distances are reduced after the code-comment synchronization, where these samples are referred to as Effective Synchronized Samples (ESSs). This metric was first proposed in CBS \cite{yang2021significance} and can be formally defined as below: 
\begin{equation}
    ESS \ Ratio = \frac{N_{ED\_diff<0}}{N}.
\end{equation}
where $N_{ED\_diff<0}$ denotes that, the number of samples whose edit distances are reduced (i.e., $edit\_distance(\hat{y}^{(k)},y^{(k)})-edit\_distance(x^{(k)},y^{(k)})<0$).

\subsection{Implementaion Details}
\label{Implementaion Details}
For the implementation of large language models (LLMs), we invoke GPT-3.5 using the constructed prompts via the OpenAI API \cite{Codecomp30:online}. Meanwhile, we deploy the Llama3 series models locally through Huggingface \cite{HuggingF94:online} and utilize vLLM \cite{vLLM58:online} to accelerate the inference process, fixing the GPU memory utilization rate to 95\%. All locally deployed LLMs are run on four NVIDIA A40 GPUs, each with 48 GB of memory.
Regarding the hyper-parameter settings, we adopt nuclear sampling \cite{holtzmancurious} for LLMs with \textit{temperature}=0.8, \textit{sampling\_number}=10, and \textit{top}-\textit{p}=0.95.
For R$^2$ComSync, we adopt Llama3-8B/70B and GPT-3.5-Turbo as the backbone models for experiments, using the same hyper-parameter settings as above.
During the retrieval phase, we retrieve the top-{2, 4, 6, 8, 10} similar demonstrations as examples (i.e., shot numbers $P$) and select the appropriate number of shots based on each model’s performance.
In the re-ranking phase, we test various combinations of $\sigma$ and $\epsilon$, and based on our exploration in Section \ref{Contributions of Multi-turn Re-ranking Strategy (RQ4)}, we finally set the thresholds for Rule 2 ($\sigma$) and Rule 3 ($\epsilon$) to 0.35 and 0.25 for Liu’s dataset, 0.35 and 0.55 for Panth’s dataset, and 0.35 and 0.2 for Pai’s dataset, respectively.

\section{Evaluation}
The following section introduces four research questions to investigate the performance of R$^2$ComSync as follows:
\begin{itemize}
    \item \textbf{RQ2:} How effective is the R$^2$ComSync compared with the state-of-the-art baselines?
    \item \textbf{RQ3:} What is the contribution of the retrieval module of R$^2$ComSync. 
    \item \textbf{RQ4:} What is the contribution of the re-ranking module of R$^2$ComSync. 
    \item \textbf{RQ5:} Does R$^2$ComSync apply to small models?
\end{itemize}

\begin{table}[htbp]
\caption{Code-Comment Synchronization Results of Each Approach}\label{Experimental Results1}

\setlength{\tabcolsep}{1.4mm}{
\begin{tabularx}{\columnwidth}{cl *{4}{>{\centering\arraybackslash}X}}
\toprule
\textbf{Dataset}& \textbf{Approach} & \textbf{Accuracy\%$^\uparrow$} & \textbf{Recall@5\%$^\uparrow$}  & \textbf{ESS Ratio\%$^\uparrow$} \\
\midrule
\multirow{9}{*}{\shortstack{Liu's\\Dataset}} 
    &CUP & $21.63_{\pm1.672}$($\textcolor{red}{5.21e\text{-}3}$) & $31.79_{\pm1.478}$($\textcolor{red}{5.96e\text{-}3}$)  & $28.64_{\pm1.512}$($\textcolor{red}{6.09e\text{-}3}$)  \\
    &HEBCUP & $26.68_{\pm0.226}$($\textcolor{red}{5.21e\text{-}3}$) & $27.67_{\pm0.226}$($\textcolor{red}{5.71e\text{-}3}$)  & $27.72_{\pm0.025}$($\textcolor{red}{3.75e\text{-}3}$)  \\
    &CBS & $28.53_{\pm0.270}$($\textcolor{red}{4.97e\text{-}3}$) & $35.76_{\pm1.338}$($\textcolor{red}{2.96e\text{-}2}$)  & $31.61_{\pm0.418}$($\textcolor{red}{5.96e\text{-}3}$)  \\
    &HatCUP & 24.30 & 35.20  & 31.33  \\
    &Toper & 30.10 & 33.04  & 34.59  \\
    \cmidrule(l{-0.05em}r{-0.05em}){2-5}
    & \multicolumn{4}{c}{\cellcolor{gray!20}\textbf{R²ComSync Scenarios}} \\
    \cmidrule(l{-0.05em}r{-0.05em}){2-5}
    &GPT3.5-turbo &$29.89_{\pm0.443}$ & $33.79_{\pm0.256}$  & \cellcolor[HTML]{E1FFFF}$36.59_{\pm0.256}$  \\
    &Llama3-8B & $26.68_{\pm0.993}$ & $30.87_{\pm0.981}$  & \cellcolor[HTML]{E1FFFF}$36.25_{\pm0.1.054}$  \\
    &Llama3-70B & \cellcolor[HTML]{E1FFFF}\textbf{\boldmath$34.08_{\pm0.133}$} & \cellcolor[HTML]{E1FFFF}\textbf{\boldmath$36.79_{\pm0.474}$}  & \cellcolor[HTML]{E1FFFF}\textbf{\boldmath$42.28_{\pm0.504}$} \\
\midrule
\multirow{7}{*}{\shortstack{Panth's\\Dataset}} 
    &CUP & $7.56_{\pm0.981}$($\textcolor{red}{5.96e\text{-}3}$) & $17.34_{\pm0.494}$($\textcolor{red}{5.96e\text{-}3}$) & $21.20_{\pm1.481}$($\textcolor{red}{5.96e\text{-}3}$)  \\
    &HEBCUP & $8.90_{\pm0.109}$($\textcolor{red}{5.71e\text{-}3}$) & $9.24_{\pm0.000}$($\textcolor{red}{3.65e\text{-}3}$)  & $19.54_{\pm0.323}$($\textcolor{red}{5.96e\text{-}3}$)  \\
    &CBS & $11.82_{\pm0.657}$($\textcolor{red}{5.83e\text{-}3}$) & $17.66_{\pm0.420}$($\textcolor{red}{5.83e\text{-}3}$)  & $25.63_{\pm0.832}$($\textcolor{red}{5.96e\text{-}3}$)  \\ 
    \cmidrule(l{-0.05em}r{-0.05em}){2-5}
    & \multicolumn{4}{c} {\cellcolor{gray!20}\textbf{R²ComSync Scenarios}} \\
    \cmidrule(l{-0.05em}r{-0.05em}){2-5}
    &GPT3.5-turbo & \cellcolor[HTML]{E1FFFF}$16.14_{\pm0.187}$ & \cellcolor[HTML]{E1FFFF}$17.99_{\pm0.187}$  & \cellcolor[HTML]{E1FFFF}$26.32_{\pm0.990}$  \\
    &Llama3-8B & \cellcolor[HTML]{E1FFFF}$12.38_{\pm1.105}$ & $15.24_{\pm1.111}$  & $21.11_{\pm1.983}$  \\
    &Llama3-70B & \cellcolor[HTML]{E1FFFF}\textbf{\boldmath$22.06_{\pm0.538}$} &  \cellcolor[HTML]{E1FFFF}\textbf{\boldmath$25.56_{\pm0.645}$}  & \cellcolor[HTML]{E1FFFF}\textbf{\boldmath$39.92_{\pm1.053}$} \\
\midrule
\multirow{7}{*}{\shortstack{Pai's\\Dataset}} 
    &CUP & $3.93_{\pm0.434}$($\textcolor{red}{5.71e\text{-}3}$) & $9.33_{\pm1.545}$($\textcolor{red}{5.96e\text{-}3}$) & $8.60_{\pm1.322}$($\textcolor{red}{5.96e\text{-}3}$)  \\
    &HEBCUP & $3.00_{\pm0.000}$($\textcolor{red}{3.54e\text{-}3}$) & $3.00_{\pm0.000}$($\textcolor{red}{3.75e\text{-}3}$)  & $6.53_{\pm0.186}$($\textcolor{red}{5.33e\text{-}3}$)  \\
    &CBS & $6.14_{\pm0.506}$($\textcolor{red}{5.58e\text{-}3}$) & $10.73_{\pm1.254}$($\textcolor{red}{6.09e\text{-}3}$)   & $12.2_{\pm1.304}$($\textcolor{red}{5.83e\text{-}3}$)  \\ 
    \cmidrule(l{-0.05em}r{-0.05em}){2-5}
    & \multicolumn{4}{c} {\cellcolor{gray!20}\textbf{R²ComSync Scenarios}} \\
    \cmidrule(l{-0.05em}r{-0.05em}){2-5}
    &GPT3.5-turbo & \cellcolor[HTML]{E1FFFF}$8.89_{\pm0.567}$ & $10.22_{\pm0.831}$  &$12.00_{\pm0.981}$  \\
    &Llama3-8B & \cellcolor[HTML]{E1FFFF}$7.80_{\pm0.909}$ & \cellcolor[HTML]{E1FFFF}$11.13_{\pm0.542}$  & \cellcolor[HTML]{E1FFFF}$14.80_{\pm1.543}$  \\
    &Llama3-70B & \cellcolor[HTML]{E1FFFF}\textbf{\boldmath$10.67_{\pm0.558}$} &  \cellcolor[HTML]{E1FFFF}\textbf{\boldmath$13.80_{\pm0.618}$}  & \cellcolor[HTML]{E1FFFF}\textbf{\boldmath$19.40_{\pm0.827}$} \\
\bottomrule
\end{tabularx}
\smallskip
\parbox{\textwidth}{\footnotesize \textit{Note:} $p$-$value$ represents the hypothetical test value between each baseline and R$^2$ComSync(Llama3-70B), the values below 0.05 are highlighted in $\textcolor{red}{red}$.}}
\end{table}

\subsection{Effectiveness of R$^2$ComSync (RQ2)}
\label{Effectiveness of R$^2$ComSync (RQ2)}
\textbf{Objective:} Various well-designed CCS approaches have continuously emerged in recent years, yet the LLM-based approach, e.g., R$^2$ComSync, to the best of our knowledge, is proposed for the first time. Therefore, we conduct the following experiments to investigate its effectiveness and make comparisons with state-of-the-art baselines.\\
\textbf{Experimental Design:} 
Experiments towards RQ2 begin with an automatic evaluation using the metrics mentioned in \ref{Evaluation Metrics}. The experimental settings are almost the same as RQ1, except we augment each LLM with our proposed R$^2$ComSync framework to make comparisons with SOTA approaches again.
Since the ultimate goal of CCS is to reduce developers' efforts in maintaining comments when code changes, we also conduct a human evaluation as a complementary experiment. Specifically, we consider three practical aspects: similarity (whether the generated new comment is similar to the reference comment), naturalness (whether the generated new comment conforms to grammar and is fluent), and sensitivity (whether the generated new comment is keenly aware of the code changes). The score for each aspect is an integer ranging from 0 to 4 (i.e., from bad to good). Similar to previous studies \cite{li2023towards,gao2020generating,hu2020deep}, for each dataset, we randomly select 50 samples from the testing set and divide them evenly into five groups. For each group, we list the generated new comments of each approach and their corresponding CCS targets in random order through questionnaires. Subsequently, we invite ten developers from the industrial community with 3-5 years of programming experience in Java and Python. We then ask them to work in pairs to anonymously carry out the evaluation, where the final scores are averaged based on each pair's results. Evaluators are allowed to search the Internet for unfamiliar concepts.\\
\textbf{Results:} Table \ref{Experimental Results1} demonstrates the CCS performance of R$^2$ComSync embedded with various LLMs. As can be seen, compared with zero-shot learning and two-shot learning across diverse datasets, LLMs obtain significant improvements, demonstrating that employing R$^2$ComSync can effectively harness the LLMs' capabilities for CCS tasks. Besides, Table \ref{Experimental Results1} also showcases the SOTA approaches' performances. Apparently, R$^2$ComSync with different LLMs demonstrates significant superiority over current SOTA approaches overall, especially in Panth and Pai's datasets. Even the worst LLM, namely Llama3-8B, surpasses almost all SOTA approaches after being equipped with the R$^2$ComSync framework. In particular, R$^2$ComSync with Llama3-70B performs the best across all datasets. For example, on Liu's dataset, R$^2$ComSync with Llama3-70B outperforms SOTA approaches by 31.64\% in terms of Accuracy, 13.16\% in Recall@5, and 37.98\% in ESS Ratio on average. Besides, it also surpasses existing approaches by 142.1\%, 89.59\%, and 82.79\% in terms of each metric on Panth's dataset. Regarding the CCS for Python programs on Pai's dataset, R$^2$ComSync with Llama3-70B still maintains its substantial superiority by 166.98\%, 145.51\%, and 127.23\% in terms of each metric in order. The above analysis demonstrates that R$^2$ComSync possesses a powerful CCS capability for the overall samples and can generate a great amount of \textit{correct synchronization}s regardless of its first attempt or five attempts. Moreover, experiments across both Java and Python projects prove that R$^2$ComSync can be extended and generalized to different programming languages. To examine the significance of the improvement, we conduct Wilcoxon Signed-Rank Tests (WSRT) \cite{wilcoxon1992individual} with a confidence level of 95\% between R$^2$ComSync(Llama3-70B) and every SOTA approach with respect to each metric, where the p-values are listed behind their respective scores under each metric. Subsequently, we further conduct Cliff's Delta analysis \cite{grissom2005effect} to measure the effect size in pairs. Results demonstrate that the superiority of R$^2$ComSync with Llama3-70B is statistically significant and consistently shows a large effect size.

\begin{table}[htbp]\caption{Human Evaluation: Code-Comment Synchronization  Results of Each Approach}\label{Human Evaluation: Code-Comment Synchronization Results of Each Approach}
\centering
\begin{tabularx}{\columnwidth}{cl *{4}{>{\centering\arraybackslash}X}}
\toprule
\textbf{Dataset} &\textbf{Approaches} & \textbf{Similarity$^\uparrow$} & \textbf{Naturalness$^\uparrow$} & \textbf{Sensitivity$^\uparrow$} \\
\midrule
\multirow{9}{*}{\shortstack{Liu's\\Dataset}} 
    &CUP & $2.11_{\pm0.11}$ $\textcolor{red}{(2.30e\text{-}5)}$ & $3.42_{\pm0.06}$ $\textcolor{red}{(3.84e\text{-}5)}$ & $1.92_{\pm0.14}$ $\textcolor{red}{(1.04e\text{-}5)}$ \\
    &HEBCUP & $2.11_{\pm0.01}$ $\textcolor{red}{(3.55e\text{-}5)}$ & $3.13_{\pm0.09}$ $\textcolor{red}{(3.87e\text{-}8)}$ & $1.99_{\pm0.17}$ $\textcolor{red}{(2.71e\text{-}5)}$ \\
    &HatCUP & $1.89_{\pm0.03}$ $\textcolor{red}{(4.46e\text{-}7)}$ & $2.75_{\pm0.13}$ $\textcolor{red}{(3.55e\text{-}15)}$ & $1.82_{\pm0.14}$ $\textcolor{red}{(1.38e\text{-}6)}$\\
    &CBS & $2.08_{\pm0.06}$ $\textcolor{red}{(1.96e\text{-}5)}$ & $3.25_{\pm0.05}$ $\textcolor{red}{(2.28e\text{-}6)}$ & $1.90_{\pm0.12}$ $\textcolor{red}{(9.08e\text{-}6)}$ \\
    &Toper & $1.74_{\pm0.02}$ $\textcolor{red}{(9.67e\text{-}8)}$ & $2.34_{\pm0.16}$ $\textcolor{red}{(1.60e\text{-}16)}$ & $1.59_{\pm0.15}$ $\textcolor{red}{(6.41e\text{-}8)}$ \\
    \cmidrule(l{-0.05em}r{-0.05em}){2-5}
    & \multicolumn{4}{c}{\cellcolor{gray!20}\textbf{R²ComSync Scenarios}} \\
    \cmidrule(l{-0.05em}r{-0.05em}){2-5}
    &Llama3-8B & \cellcolor[HTML]{E1FFFF} $2.34_{\pm0.10}$ & \cellcolor[HTML]{E1FFFF} $3.36_{\pm0.04}$ & \cellcolor[HTML]{E1FFFF} $2.26_{\pm0.14}$ \\
    &GPT3.5-turbo & \cellcolor[HTML]{E1FFFF} $2.90_{\pm0.08}$ & \cellcolor[HTML]{E1FFFF} $3.84_{\pm0.08}$ & \cellcolor[HTML]{E1FFFF} $2.84_{\pm0.12}$ \\
    &Llama3-70B & \cellcolor[HTML]{E1FFFF} \textbf{\boldmath$3.04_{\pm0.04}$} & \cellcolor[HTML]{E1FFFF}\textbf{\boldmath$3.85_{\pm0.05}$} & \cellcolor[HTML]{E1FFFF}\textbf{\boldmath$2.99_{\pm0.05}$} \\
\midrule
\multirow{7}{*}{\shortstack{Panth's\\Dataset}} 
    &CUP & $1.82_{\pm0.18}$ $\textcolor{red}{(1.37e\text{-}3)}$ & $3.62_{\pm0.10}$ $\textcolor{red}{(3.86e\text{-}2)}$ & $1.60_{\pm0.26}$ $\textcolor{red}{(2.03e\text{-}4)}$ \\
    &HEBCUP & $1.82_{\pm0.16}$ $\textcolor{red}{(1.72e\text{-}3)}$ & $3.28_{\pm0.04}$ $\textcolor{red}{(6.70e\text{-}4)}$ & $1.58_{\pm0.26}$ $\textcolor{red}{(3.76e\text{-}4)}$ \\
    &CBS & $2.13_{\pm0.15}$ $\textcolor{red}{(3.62e\text{-}2)}$ & $3.56_{\pm0.08}$ $\textcolor{red}{(2.20e\text{-}2)}$ & $1.95_{\pm0.21}$ $\textcolor{red}{(1.10e\text{-}2)}$ \\
    \cmidrule(l{-0.05em}r{-0.05em}){2-5}
    & \multicolumn{4}{c}{\cellcolor{gray!20}\textbf{R²ComSync Scenarios}} \\
    \cmidrule(l{-0.05em}r{-0.05em}){2-5}
    &Llama3-8B & \cellcolor[HTML]{E1FFFF}$1.86_{\pm0.14}$& $3.06_{\pm0.00}$ & \cellcolor[HTML]{E1FFFF}$1.96_{\pm0.16}$ \\
    &GPT3.5-turbo & \cellcolor[HTML]{E1FFFF}\textbf{\boldmath$2.87_{\pm0.03}$}  & \cellcolor[HTML]{E1FFFF}\textbf{\boldmath$3.79_{0.01}$} & \cellcolor[HTML]{E1FFFF}\textbf{\boldmath$2.97_{\pm0.05}$} \\
    &Llama3-70B & \cellcolor[HTML]{E1FFFF}$2.66_{\pm0.04}$ & \cellcolor[HTML]{E1FFFF}\textbf{\boldmath$3.79_{\pm0.03}$} & \cellcolor[HTML]{E1FFFF}$2.67_{\pm0.17}$ \\
\midrule
\multirow{7}{*}{\shortstack{Pai's\\Dataset}} 
    &CUP & $1.88_{\pm0.13}$ $\textcolor{red}{(1.39e\text{-}12)}$ & $2.37_{\pm0.15}$ $\textcolor{red}{(3.12e\text{-}15)}$ & $0.79_{\pm0.33}$ $\textcolor{red}{(7.13e\text{-}14)}$ \\
    &HEBCUP & $2.31_{\pm0.13}$ $\textcolor{red}{(9.19e\text{-}11)}$ & $2.81_{\pm0.19}$ $\textcolor{red}{(2.19e\text{-}14)}$ & $1.19_{\pm0.33}$ $\textcolor{red}{(3.49e\text{-}12)}$ \\
    &CBS & $2.12_{\pm0.24}$ $\textcolor{red}{(1.22e\text{-}11)}$ & $2.78_{\pm0.24}$ $\textcolor{red}{(1.22e\text{-}14)}$ & $1.04_{\pm0.14}$ $\textcolor{red}{(6.70e\text{-}14)}$ \\
    \cmidrule(l{-0.05em}r{-0.05em}){2-5}
    & \multicolumn{4}{c}{\cellcolor{gray!20}\textbf{R²ComSync Scenarios}} \\
    \cmidrule(l{-0.05em}r{-0.05em}){2-5}
    &Llama3-8B & \cellcolor[HTML]{E1FFFF}$2.84_{\pm0.12}$& \cellcolor[HTML]{E1FFFF}$3.32_{\pm0.12}$ & \cellcolor[HTML]{E1FFFF}$2.47_{\pm0.11}$ \\
    &GPT3.5-turbo & \cellcolor[HTML]{E1FFFF}\textbf{\boldmath$3.44_{\pm0.06}$}& \cellcolor[HTML]{E1FFFF}\textbf{\boldmath$3.83_{\pm0.07}$} & \cellcolor[HTML]{E1FFFF}\textbf{\boldmath$3.19_{\pm0.01}$}\\
    &Llama3-70B & \cellcolor[HTML]{E1FFFF}$3.32_{\pm0.12}$ & \cellcolor[HTML]{E1FFFF}$3.74_{\pm0.12}$ & \cellcolor[HTML]{E1FFFF}$3.05_{\pm0.09}$ \\
\bottomrule
\end{tabularx}
\end{table}

Furthermore, Table \ref{Human Evaluation: Code-Comment Synchronization Results of Each Approach} demonstrates the human evaluation results. To validate the effectiveness of the assessment, we compute Cohen’s kappa coefficient \cite{chmura2002kappa} of agreement for the
scores assigned by the two evaluators, finding that it consistently exceeds 0.55 across all evaluation metrics and datasets. Therefore, the evaluation process is above moderate consistency.
As can be seen, compared with the results of automatic evaluation (Figure \ref{Experimental Results1}), R$^2$ComSync demonstrates a more consistent superiority against SOTA approaches. Even using a relatively less capable model such as Llama3-8B, it still achieves higher scores than SOTA baselines in most scenarios. Taking the best-performing LLM, i.e., Llama3-70B, and SOTA approaches to conduct WSRT among 50 pairs of evaluation results on each dataset, the hypothetical tests concerning the improvements are all statistically significant (p-value$<$0.05) as shown in Table \ref{Human Evaluation: Code-Comment Synchronization Results of Each Approach}. Moreover, the Cliff's Delta analysis also reveals that their effect sizes are all non-negligible.
This proves the practical value of R$^2$ComSync, showing that R$^2$ComSync can generate comments that developers prefer, not just in terms of correctness (i.e., similarity and sensitivity), but also in terms of naturalness. We attribute the outperformance in human evaluation to using LLMs as R$^2$ComSync's backbone, as they were pretrained on massive human-written corpora, providing a solid foundation for generating human-preference text.
Another worth noting phenomenon is that GPT-3.5-Turbo has closed the distance with Llama3-70B when equipped with R$^2$ComSync in the human evaluation. Particularly, in Panth and Pai's datasets, it even slightly surpasses Llama3-70B across all evaluation metrics. Although the WSRT has revealed that the performance differences between them are not significant, it still illustrates that R$^2$ComSync with GPT-3.5-Turbo tends to generate new comments that are more satisfying to human beings than similar to ground truths.


\begin{boxK}
 \faForward \textbf{RQ2:} R$^2$ComSync significantly lifts LLMs' performance in the CCS task, and most studied LLMs outperform SOTA approaches after equipping with R$^2$ComSync, especially in the human evaluations.
\end{boxK}

\begin{table}[!htbp]\caption{Experimental Results Concerning the Influence of EHR}\label{Experimental Results Concerning the Influence of EHR}
\setlength{\tabcolsep}{1.4mm}{
\begin{tabularx}{\columnwidth}{cl *{4}{>{\arraybackslash}X}}
\toprule
\textbf{Dataset}&\textbf{Models} & \textbf{Methods} & \textbf{Accuracy\%$^\uparrow$} & \textbf{Recall@5\%$^\uparrow$} & \textbf{ESS Ratio\%$^\uparrow$} \\
\midrule
\multirow{13}{*}{\shortstack{Liu's\\Dataset}}
    &\multirow{4}{*}{GPT3.5-turbo}
        & Random & $18.61_{\pm0.287}$ & $26.82_{\pm0.304}$ & $24.85_{\pm0.469}$ \\
        && Expert & $20.21_{\pm1.793}$ & $27.81_{\pm1.112}$ & $26.38_{\pm1.275}$ \\
        && CodeBERT & \underline{$23.78_{\pm0.501}$} & \underline{$31.16_{\pm0.392}$} & \underline{$29.38_{\pm1.109}$} \\
        && EHR & \textbf{\boldmath$25.78_{\pm0.720}$} & \textbf{\boldmath$32.07_{\pm0.390}$} & \textbf{\boldmath$31.54_{\pm0.650}$} \\
    \cmidrule(lr){2-6}
    &\multirow{4}{*}{Llama3-8B}
        & Random & $7.45_{\pm0.550}$ & $15.38_{\pm0.594}$ & $10.22_{\pm0.706}$ \\
        && Expert & $11.36_{\pm0.827}$ & $19.43_{\pm0.863}$ & $15.74_{\pm1.132}$ \\
        && CodeBERT & \textbf{\boldmath$16.26_{\pm0.703}$} & \textbf{\boldmath$25.48_{\pm0.699}$} & \textbf{\boldmath$21.02_{\pm0.877}$} \\
        && EHR & \underline{$15.62_{\pm0.922}$} & \underline{$24.67_{\pm0.809}$} & \underline{$20.09_{\pm1.033}$} \\
    \cmidrule(lr){2-6}
    &\multirow{4}{*}{Llama3-70B}
        & Random & $25.54_{\pm0.805}$ & $31.03_{\pm0.301}$ & $32.74_{\pm1.021}$ \\
        && Expert & $27.67_{\pm0.605}$ & $32.83_{\pm0.435}$ & $34.97_{\pm0.807}$ \\
        && CodeBERT & \underline{$29.37_{\pm1.060}$} & \underline{$34.32_{\pm0.828}$} & \underline{$37.67_{\pm1.346}$} \\
        && EHR & \textbf{\boldmath$30.14_{\pm0.609}$} & \textbf{\boldmath$34.77_{\pm0.507}$} & \textbf{\boldmath$38.37_{\pm0.876}$} \\
\midrule
\multirow{13}{*}{\shortstack{Panth's\\Dataset}}
    &\multirow{4}{*}{GPT3.5-turbo}
        & Random & $10.61_{\pm0.470}$ & $15.53_{\pm0.470}$ & $18.19_{\pm0.470}$ \\
        && Expert &\textbf{\boldmath$13.99_{\pm0.737}$} & \textbf{\boldmath$18.84_{\pm0.518}$} & \textbf{\boldmath$23.02_{\pm1.002}$} \\
        && CodeBERT & $11.51_{\pm0.494}$ & $16.24_{\pm0.494}$ & $20.00_{\pm0.626}$ \\
        && EHR & \underline{$12.59_{\pm0.614}$} & \underline{$16.75_{\pm0.415}$} & \underline{$20.56_{\pm1.282}$} \\
    \cmidrule(lr){2-6}
    &\multirow{4}{*}{Llama3-8B}
        & Random & $2.04_{\pm0.214}$ & $5.95_{\pm1.022}$ & $9.23_{\pm0.816}$ \\
        && Expert & \textbf{\boldmath$5.35_{\pm0.871}$} & \textbf{\boldmath$10.92_{\pm0.872}$} & \textbf{\boldmath$13.05_{\pm1.206}$} \\
        && CodeBERT & $4.32_{\pm0.738}$ & $9.40_{\pm0.760}$ & $12.29_{\pm1.258}$ \\
        && EHR & \underline{$4.35_{\pm0.693}$} & \underline{$9.83_{\pm0.775}$} & \underline{$12.48_{\pm0.999}$} \\
    \cmidrule(lr){2-6}
    &\multirow{4}{*}{Llama3-70B}
        & Random & $15.66_{\pm1.183}$ & $20.51_{\pm0.682}$ & $28.36_{\pm1.366}$ \\
        && Expert & \textbf{\boldmath$19.40_{\pm0.717}$} & \textbf{\boldmath$24.32_{\pm0.549}$} & \underline{$33.13_{\pm1.046}$} \\
        && CodeBERT & $16.94_{\pm0.862}$ & $22.16_{\pm0.626}$& $30.63_{\pm1.456}$ \\
        && EHR & \underline{$18.11_{\pm0.717}$} & \underline{$23.48_{\pm0.549}$} & \textbf{\boldmath$33.14_{\pm1.046}$} \\
\midrule
\multirow{13}{*}{\shortstack{Pai's\\Dataset}}
    &\multirow{4}{*}{GPT3.5-turbo}
        & Random & $5.53_{\pm0.272}$ & $8.60_{\pm0.272}$ & $10.53_{\pm0.272}$ \\
        && Expert &$5.69_{\pm0.470}$ & $9.20_{\pm0.496}$ & \underline{$12.20_{\pm0.565}$} \\
        && CodeBERT & \textbf{\boldmath$6.73_{\pm0.758}$} & \textbf{\boldmath$9.84_{\pm{0.396}}$} & $11.58_{\pm0.587}$ \\
        && EHR & \underline{$6.38_{\pm0.360}$} & \underline{$9.56_{\pm0.466}$} & \textbf{\boldmath$12.31_{\pm0.981}$} \\
    \cmidrule(lr){2-6}
    &\multirow{4}{*}{Llama3-8B}
        & Random & $1.78_{\pm0.447}$ & $4.73_{\pm0.517}$ & $9.40_{\pm0.796}$ \\
        && Expert & $2.29_{\pm0.630}$ & $5.76_{\pm0.668}$ & $9.59_{\pm0.673}$ \\
        && CodeBERT & \textbf{\boldmath$3.24_{\pm0.776}$} & \underline{$7.25_{\pm0.731}$} & \textbf{\boldmath$11.60_{\pm0.884}$} \\
        && EHR & \underline{$3.23_{\pm0.814}$} & \textbf{\boldmath$7.33_{\pm0.724}$} & \underline{$10.81_{\pm1.538}$} \\
    \cmidrule(lr){2-6}
    &\multirow{4}{*}{Llama3-70B}
        & Random & $4.76_{\pm0.666}$ & $9.31_{\pm0.380}$ & $14.18_{\pm1.118}$ \\
        && Expert & \underline{$5.93_{\pm0.578}$} & \underline{$9.88_{\pm0.434}$} & $14.97_{\pm1.109}$ \\
        && CodeBERT & $5.55_{\pm0.435}$ & $9.20_{\pm0.640}$& \underline{$16.67_{\pm0.872}$} \\
        && EHR & \textbf{\boldmath$7.45_{\pm0.636}$} & \textbf{\boldmath$11.69_{\pm0.553}$} & \textbf{\boldmath$17.61_{\pm0.940}$} \\
\bottomrule
 \end{tabularx}}
\smallskip
\parbox{\textwidth}{\footnotesize \textit{Note:}  Each model's best-performing results are highlighted in \textbf{bold}, and their second-best-performing results are highlighted with \underline{underlines}.}
\end{table}

\subsection{Contributions of Ensemble Hybrid Retrieval (RQ3)}
\label{Contributions of Ensemble Hybrid Retrieval (RQ3)}
\textbf{Objective:} Although we have verified the effectiveness of R$^2$ComSync as a whole in Section \ref{Effectiveness of R$^2$ComSync (RQ2)}, we still need to quantify the contribution of the Ensemble Hybrid Retrieval (EHR) method, thereby proving our motivation for retrieving CCS demonstrations from perspectives of both code-comment semantic and change patterns is worthwhile.\\
\textbf{Experimental Design:} 
To carry out the ablation study for the EHR module, we first remove the MR strategy to present the performance of our proposal equipped with EHR solely. Since the EHR module is a composite method of CodeBERT retrieval and expert retrieval, we compare the EHR module with the above two retrieval methods, respectively. Besides, we also include the random selection of demonstrations for comparison. To eliminate the performance discrepancies across different shot numbers ($P$), we feed LLMs with $P=\{2, 4, 6, 8, 10\}$ in order and then average their performances to evaluate the contribution of the EHR module and make comparisons with other retrieval methods. It should be noted that experiments are repeated five times for each setting of $P$.
Subsequently, based on the above experimental results with different shot numbers, we investigate the performance variation. All the above experiments are conducted across all datasets with all LLMs under test.
\\
\textbf{Results:} 
First, we compared the impact of different retrieval methods on LLMs' performance, where the specific experimental results are shown in Table \ref{Experimental Results Concerning the Influence of EHR}. As can be seen, in Liu and Pai's datasets, EHR performs the best across almost three studied LLMs on most occasions. On average, EHR outperforms CodeBERT retrieval by 9.58\% in terms of Accuracy, by 8.44\% in terms of Recall@5, and by 1.72\% in terms of ESS Ratio across all studied LLMs on Pai's dataset. As for expert retrieval, EHR still maintains its superiority by 26.27\%, 16.50\%, and 10.42\% in each metric, respectively. When it turns to random retrieval, EHR manifests much more pronounced advantages by 56.51\%, 30.56\%, and 17.70\% on each metric in order. Similar trends occur in Liu's dataset, where EHR dominates among all retrieval methods, CodeBERT retrieval performs the second best, and expert retrieval surpasses the random one.
We attribute this to the fact that CCS samples of Liu and Pai's dataset are relatively longer (See Section \ref{Dataset} for details), especially Pai's dataset, and expert features solely may have difficulties in fully capturing their diverse characteristics. Besides, the longest code and comment length also induce the absolute worst performance on Pai's dataset across all studied LLMs.

As for the experimental results on Panth's dataset, the expert retrieval method achieves the best performance, outperforming other retrieval methods, including EHR, by 13.74\%--72.66\% in terms of Accuracy, by 9.05\%--41.14\% in terms of Recall@5, and by 5.50\%--28.25\% in terms of ESS Ratio across all studied LLMs on average. A possible explanation is as follows: The sample distribution of Panth's dataset exhibits significant discrepancies, as shown in Table \ref{tab:dataset_stats}, with a much higher standard deviation on both code and comment lengths, as well as their corresponding NCSs. Thus, it can be regarded as a ``high-variability dataset''. CodeBERT relies on alignment between the data distribution and its pretraining data in semantic similarity tasks \cite{9609166}. A high standard deviation may disrupt this alignment, leading to performance degradation and resulting in lower-quality retrieval samples for EHR, thereby affecting its performance. 
On the contrary, expert features are more precise in sample representation for shorter ones, as these patterns are easier and explicit. Besides, they need not proactively learn crucial features among samples, leading to their surpassing in Panth's dataset.
Even though expert retrieval performs the best in Panth's dataset, our proposed EHR consistently performs second-best across almost all LLMs and evaluation metrics, demonstrating its effectiveness.
Although CCS samples in diverse datasets exhibit varying proneness to different retrieval methods, they all perform worst with random retrieval, demonstrating that relevant examples are crucial for LLM-based code-comment synchronization.
For practitioners, it may be beneficial to consider different mixing ratios for EHR depending on the degree of dispersion in the sample distribution. For datasets with more dispersed sample distributions or shorter samples, a larger weight can be assigned to expert retrieval; otherwise, a larger weight can be assigned to CodeBERT's retrieval results.

\begin{figure}
    \centering
    \includegraphics[width=1\linewidth]{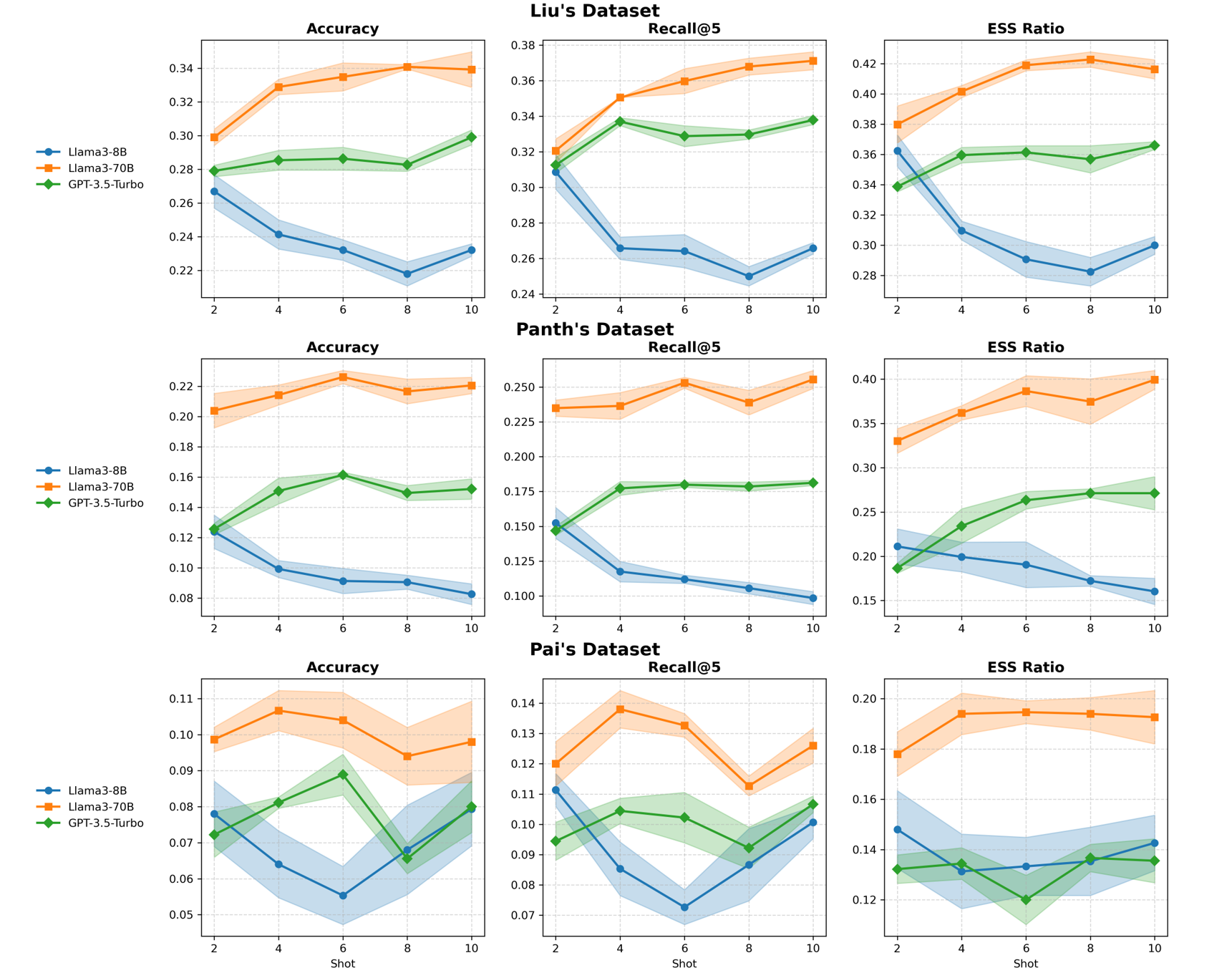}
    \caption{LLMs' Performance Variation among Different Shot Numbers across three different datasets.}
    \label{mertricVSshot}
\end{figure}
Next, based on the above experimental results, we further analyze the impact of different shot numbers ($P$) on the LLMs' performance, where the experimental results are shown in Figure \ref{mertricVSshot}.
As can be seen, for GPT-3.5-Turbo and Llama3-70B, with more CCS demonstrations, they tend to perform gradually better, but this tendency cannot last forever; it drops at a certain shot number, which is also the same in other in-context learning scenarios \cite{xue2024automated,li2023towards,dong2022survey}. For Liu and Panth's dataset, their peak values tend to be lagged compared to those in Pai's dataset, as the latter has much longer code and comment lengths on average. This makes it harder for the above LLMs to leverage the relatively more complex contextual information, given the same shot numbers. Nonetheless, as for Llama3-8B, it performs a declining trend overall, regardless of any CCS dataset. Owing to the fact that LLMs' performance scales following a power law with respect to parameter count, data volume, and computational resources, enabling more effective utilization of additional contextual information \cite{kaplan2020scaling}. Hence, Llama3-8B apparently has difficulty in harnessing the given contextual information as the number of shots increases, due to its smaller parameter size.
Therefore, selecting an appropriate shot number based on diverse LLMs and datasets can enhance the performance of code-comment synchronization.

\begin{boxK}
 \faForward \textbf{RQ3:} The EHR method substantially enhances LLMs' performance on the CCS task and either performs the best or second best among different retrieval methods. Moreover, for different LLMs and datasets, selecting an appropriate shot number is also crucial.
\end{boxK}

\begin{table}[!htbp]\caption{Results Concerning the Influence of Multi-turn Re-ranking}
\label{the Influence of Multi-turn Re-ranking}
\setlength{\tabcolsep}{1.4mm}{
\begin{tabularx}{\columnwidth}{cl *{5}{>{\arraybackslash}X}}
\toprule
\textbf{Dataset} & \textbf{Model} & \textbf{Components} & \textbf{Accuracy\%$^\uparrow$} & \textbf{Recall@5\%$^\uparrow$} & \textbf{ESS Ratio\%$^\uparrow$}  \\
\midrule
\multirow{13}{*}{\shortstack{Liu's\\Dataset}} 
    &\multirow{4}{*}{GPT-3.5-Turbo}
        & W/O MR & $27.45_{\pm0.587}$ & $33.24_{\pm0.558}$  & $33.61_{\pm0.462}$  \\
        && MR-1 & $28.08_{\pm0.779}$ & $33.24_{\pm0.558}$  & $34.24_{\pm0.444}$  \\
        && MR-1-2 & $29.17_{\pm0.462}$ & $33.51_{\pm0.339}$  & $36.14_{\pm0.222}$  \\
        && MR-1-2-3 & \textbf{\boldmath$29.89_{\pm0.444}$} & \textbf{\boldmath$33.79_{\pm0.256}$}  &\textbf{\boldmath$36.59_{\pm0.256}$} \\
    \cmidrule(lr){2-6}
    & \multirow{4}{*}{Llama3-8B}
        & W/O MR & $19.95_{\pm0.966}$ & $29.46_{\pm0.935}$  & $26.96_{\pm1.347}$  \\
        && MR-1 & $20.65_{\pm0.749}$ & $29.51_{\pm0.919}$  & $27.61_{\pm1.095}$  \\
        && MR-1-2 & $24.46_{\pm1.100}$ & $30.82_{\pm0.817}$  & $34.18_{\pm1.713}$  \\
        && MR-1-2-3 & \textbf{\boldmath$26.68_{\pm0.993}$} & \textbf{\boldmath$30.87_{\pm0.981}$}  &\textbf{\boldmath$36.15_{\pm1.054}$} \\
    \cmidrule(lr){2-6}
    &\multirow{4}{*}{Llama3-70B}
        & W/O MR & $30.60_{\pm0.504}$ & $35.92_{\pm0.554}$  & $38.70_{\pm0.741}$  \\
        && MR-1 & $30.71_{\pm0.570}$ & $35.87_{\pm0.595}$  & $38.64_{\pm0.916}$  \\
        && MR-1-2 & $32.17_{\pm0.657}$ & $36.47_{\pm0.605}$  & $40.76_{\pm0.666}$  \\
        && MR-1-2-3 & \textbf{\boldmath$34.08_{\pm0.133}$} & \textbf{\boldmath$36.79_{\pm0.474}$}  &\textbf{\boldmath$42.28_{\pm0.504}$} \\
\midrule
\multirow{13}{*}{\shortstack{Panth's\\Dataset}}
    &\multirow{4}{*}{GPT-3.5-Turbo}
        & W/O MR & $13.36_{\pm0.374}$ & $17.59_{\pm0.374}$  & $22.22_{\pm1.620}$  \\
        && MR-1 & $13.36_{\pm0.374}$ & $17.59_{\pm0.374}$  & $22.22_{\pm1.620}$  \\
        && MR-1-2 & $15.21_{\pm0.815}$ & $17.86_{\pm0.324}$  & $24.74_{\pm0.935}$  \\
        && MR-1-2-3 & \textbf{\boldmath$16.14_{\pm0.187}$} & \textbf{\boldmath$17.99_{\pm0.187}$}  &\textbf{\boldmath$26.32_{\pm0.990}$} \\
    \cmidrule(lr){2-6}
    &\multirow{4}{*}{Llama3-8B}
        & W/O MR & $7.70_{\pm0.692}$ & $13.65_{\pm0.736}$  & $16.90_{\pm1.193}$  \\
        && MR-1 & $7.70_{\pm0.692}$ & $13.65_{\pm0.736}$  & $16.90_{\pm1.193}$  \\
        && MR-1-2 & $11.59_{\pm1.077}$ & $15.24_{\pm1.111}$  & $21.03_{\pm2.085}$  \\
        && MR-1-2-3 & \textbf{\boldmath$12.38_{\pm1.105}$} & \textbf{\boldmath$15.24_{\pm1.111}$}  &\textbf{\boldmath$21.11_{\pm1.983}$} \\
    \cmidrule(lr){2-6}
    &\multirow{4}{*}{Llama3-70B}
        & W/O MR & $18.25_{\pm0.905}$ & $24.29_{\pm0.985}$  & $33.97_{\pm1.397}$ \\
        && MR-1 & $18.25_{\pm0.905}$ & $23.49_{\pm0.985}$  & $34.52_{\pm1.397}$  \\
        && MR-1-2 & $20.79_{\pm0.855}$ &$23.89_{\pm0.952}$  & $36.98_{\pm2.236}$  \\
        && MR-1-2-3 & \textbf{\boldmath$22.06_{\pm0.817}$} & \textbf{\boldmath$25.56_{\pm0.884}$} &\textbf{\boldmath$39.92_{\pm2.567}$} \\
        \midrule
\multirow{13}{*}{\shortstack{Pai's\\Dataset}}
    &\multirow{4}{*}{GPT-3.5-Turbo}
        & W/O MR & $6.33_{\pm0.272}$ & $9.89_{\pm0.567}$  & $11.78_{\pm1.110}$  \\
        && MR-1 & $6.33_{\pm0.272}$ & $9.89_{\pm0.567}$  & $11.78_{\pm1.100}$  \\
        && MR-1-2 & $7.56_{\pm0.685}$ & $10.11_{\pm0.875}$  & \textbf{\boldmath$12.22_{\pm0.875}$}  \\
        && MR-1-2-3 & \textbf{\boldmath$8.89_{\pm0.567}$} & \textbf{\boldmath$10.22_{\pm0.831}$}  &$12.00_{\pm0.981}$ \\
    \cmidrule(lr){2-6}
    &\multirow{4}{*}{Llama3-8B}
        & W/O MR & $4.40_{\pm0.800}$ & $9.40_{\pm0.389}$  & $12.33_{\pm0.023}$  \\
        && MR-1 & $4.40_{\pm0.800}$ & $9.40_{\pm0.389}$  & $12.33_{\pm2.280}$  \\
        && MR-1-2 & $6.13_{\pm0.686}$ & $10.40_{\pm0.611}$  & $13.87_{\pm1.916}$  \\
        && MR-1-2-3 & \textbf{\boldmath$7.80_{\pm0.909}$} & \textbf{\boldmath$11.13_{\pm0.542}$}  &\textbf{\boldmath$14.80_{\pm1.543}$} \\
    \cmidrule(lr){2-6}
    &\multirow{4}{*}{Llama3-70B}
        & W/O MR & $8.73_{\pm0.646}$ & $12.80_{\pm0.452}$  & $18.67_{\pm0.869}$  \\
        && MR-1 & $8.73_{\pm0.646}$ & $12.80_{\pm0.452}$  & $18.67_{\pm0.869}$ \\
        && MR-1-2 & $9.80_{\pm0.581}$ &$13.40_{\pm0.646}$  & $19.33_{\pm0.471}$  \\
        && MR-1-2-3 & \textbf{\boldmath$10.67_{\pm0.558}$} & \textbf{\boldmath$13.80_{\pm0.618}$}  &\textbf{\boldmath$19.40_{\pm0.827}$} \\
    \bottomrule
\end{tabularx}}
\end{table}

\subsection{Contributions of Multi-turn Re-ranking Strategy (RQ4)}
\label{Contributions of Multi-turn Re-ranking Strategy (RQ4)}
\textbf{Objective:} Similar to RQ3, to examine whether the Multi-turn Re-ranking (MR) strategy can effectively prioritize correct-prone candidates, this subsection quantifies its contribution in R$^2$ComSync. Besides, we also analyze the threshold settings of Rules 2 (i.e., $\sigma$) and 3 (i.e., $\epsilon$) in the MR strategy, thereby investigating its influence even further. \\
\textbf{Experimental Design:} To verify the effectiveness of the MR strategy, we carry out another group of ablation experiments. Specifically, we first list R$^2$ComSync without MR strategy as the baseline (i.e., W/O MR), then adopt Rule 1 to re-rank candidates generated by LLMs (i.e., MR-1), thereby quantifying the contribution of Rule 1. Afterwards, we further impose Rules 2 (i.e., MR-1-2) and 3 (i.e., MR-1-2-3) on it successively to complete the ablation experiment. 
For different models and datasets, we refer to Figure \ref{mertricVSshot} to select the shot with the best performance for re-ranking experiments. Since the experiments for each shot number are repeated five times, their follow-up re-ranking results are also recorded, respectively, as shown in Table \ref{the Influence of Multi-turn Re-ranking}.
Regarding the study on threshold settings for Rule 2 ($\sigma$) and 3 ($\epsilon$), we investigate the performance of Llama3-70B using various threshold combinations across the three datasets.
Specifically, based on the findings in Section \ref{Motivation}, we adjust ($\sigma$) from 0.2 to 0.4 and ($\epsilon$) from 0.2 to 0.6 with a step size of 0.05.
Accordingly, we exhaustively carry out 5*9*3=135 groups of experiments to observe the influence of each pair of threshold combinations on R$^2$ComSync.\\
\textbf{Results:} Table \ref{the Influence of Multi-turn Re-ranking} demonstrates the results of ablation experiments concerning the MR strategy. Firstly, compared with R$^2$ComSync without MR strategy (i.e., W/O MR), using the rules we proposed consecutively to re-rank the candidate values can lead to significant improvements by 15.71\%-46.65\% in terms of Accuracy, 3.48\%-9.85\% in terms of Recall@5, and 8.60\%-20.29\% in terms of ESS Ratio across three datasets, indicating that re-ranking the candidates generated by LLM is necessary for the CCS field. Besides, as can be seen, both Rule 2 and Rule 3 can almost progressively boost CCS performance, proving that the effectiveness of these rules, which were summarized from Liu's datasets, can be generalized to other CCS scenarios in practice. However, Rule 1 seems to be only effective in Liu's dataset, demonstrating its overfitting to our analyzed samples from Liu's dataset. Besides, Rule 1's contribution is relatively limited compared to Rules 2 and 3, further indicating its weak level in the MR strategy.  

\begin{figure*}[htbp]
\centering
\includegraphics[width=1\textwidth]
{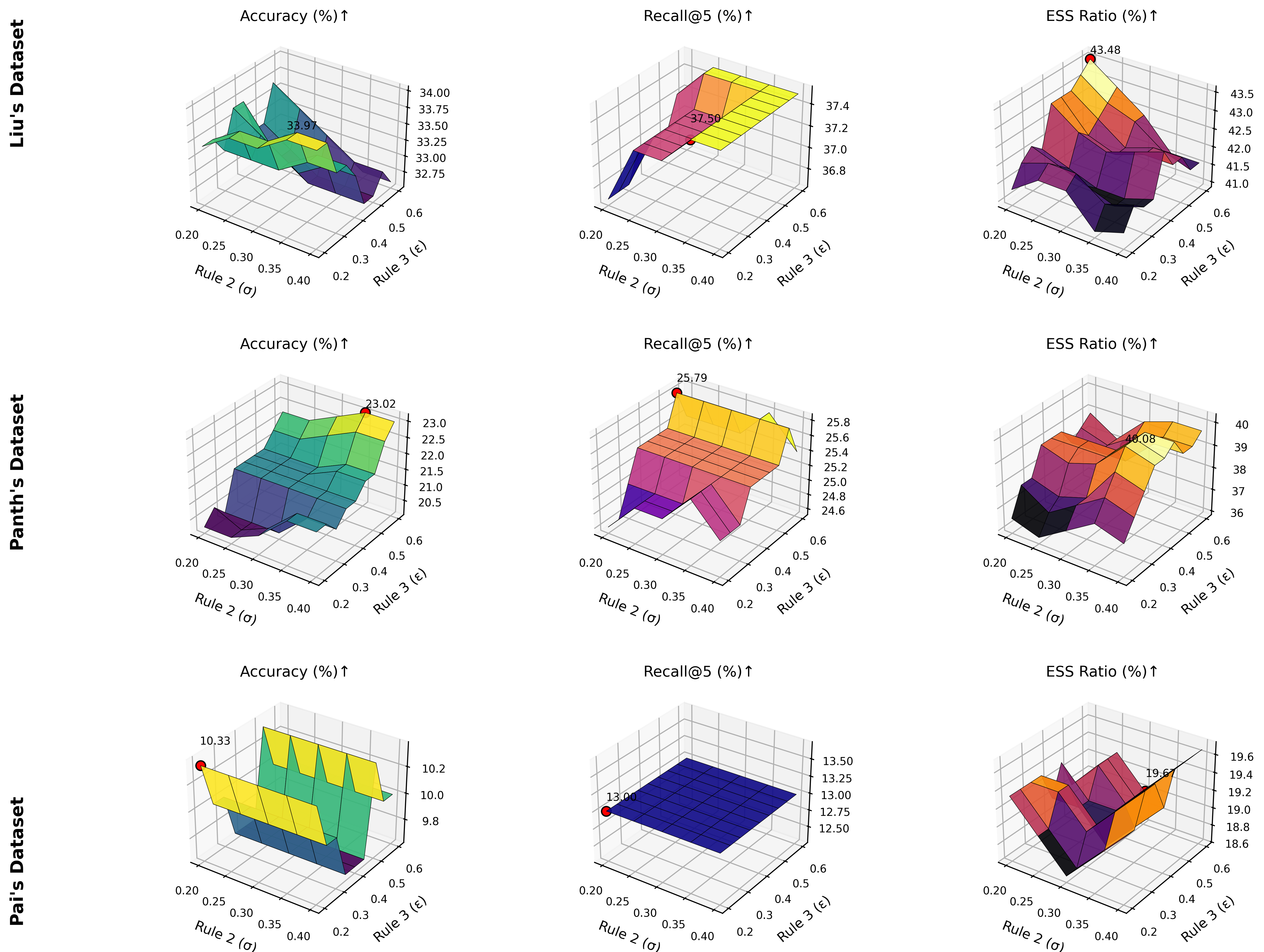}
\caption{Sensitivity Analysis of Llama3-70B’s Performance Metrics on Rule 2 ($\sigma$) and Rule 3 ($\epsilon$).}
\label{threshold}
\end{figure*}

Afterwards, we further analyze the threshold setting in Rules 2 and 3, where the experimental results are shown in Figure \ref{threshold}. 
It can be observed that on Liu's dataset, as $\sigma$ and $\epsilon$ increase, the performance of R$^2$ComSync in terms of Accuracy generally decreases, while Recall@5 and ESS Ratio initially increase and then decline. In contrast, on Panth's dataset, all three metrics exhibit an upward trend as both thresholds increase. Nonetheless, as for Pai's dataset, R$^2$ComSync's both Accuracy and ESS Ratio scores manifest a trend of declining first, then increasing, but the former is mainly controlled by $\epsilon$ while the latter is almost solely affected by $\sigma$. An interesting finding is on Recall@5, where it is consistently stable across diverse combinations of $\epsilon$ and $\sigma$. A potential explanation is that Pai's dataset has the relatively longest comments, which leads to the extreme difficulty in generating exactly \textit{correct synchronization}, given the limited attempts. Hence, even though we continuously change their candidate orders by altering re-ranking hyper-parameters, the Recall@5 keep almost unchanged. The thresholds used in previous RQs are marked with red dots in Figure \ref{threshold}, and these settings have achieved or are close to optimal. It is important to note that adjustments for $\sigma$ and $\epsilon$ will significantly affect the CCS performance of R$^2$ComSync in most scenarios. Therefore, we recommend that practitioners use our settings as a baseline and further tune the values of $\sigma$ and $\epsilon$ according to the specific circumstances to achieve better performances.

\begin{boxK}
 \faForward \textbf{RQ4:} The MR strategy is effective in prioritizing correct-prone candidates. Towards the threshold configuration, we recommend practitioners use our setting as an initial configuration and tune it further according to their specific demands.
\end{boxK}

\subsection{Applicability of R$^2$ComSync to small models (RQ5)}
\label{Applicability of R$^2$ComSync to small models (RQ5)}
\begin{figure*}[htbp]
\centering
\includegraphics[width=1\textwidth]
{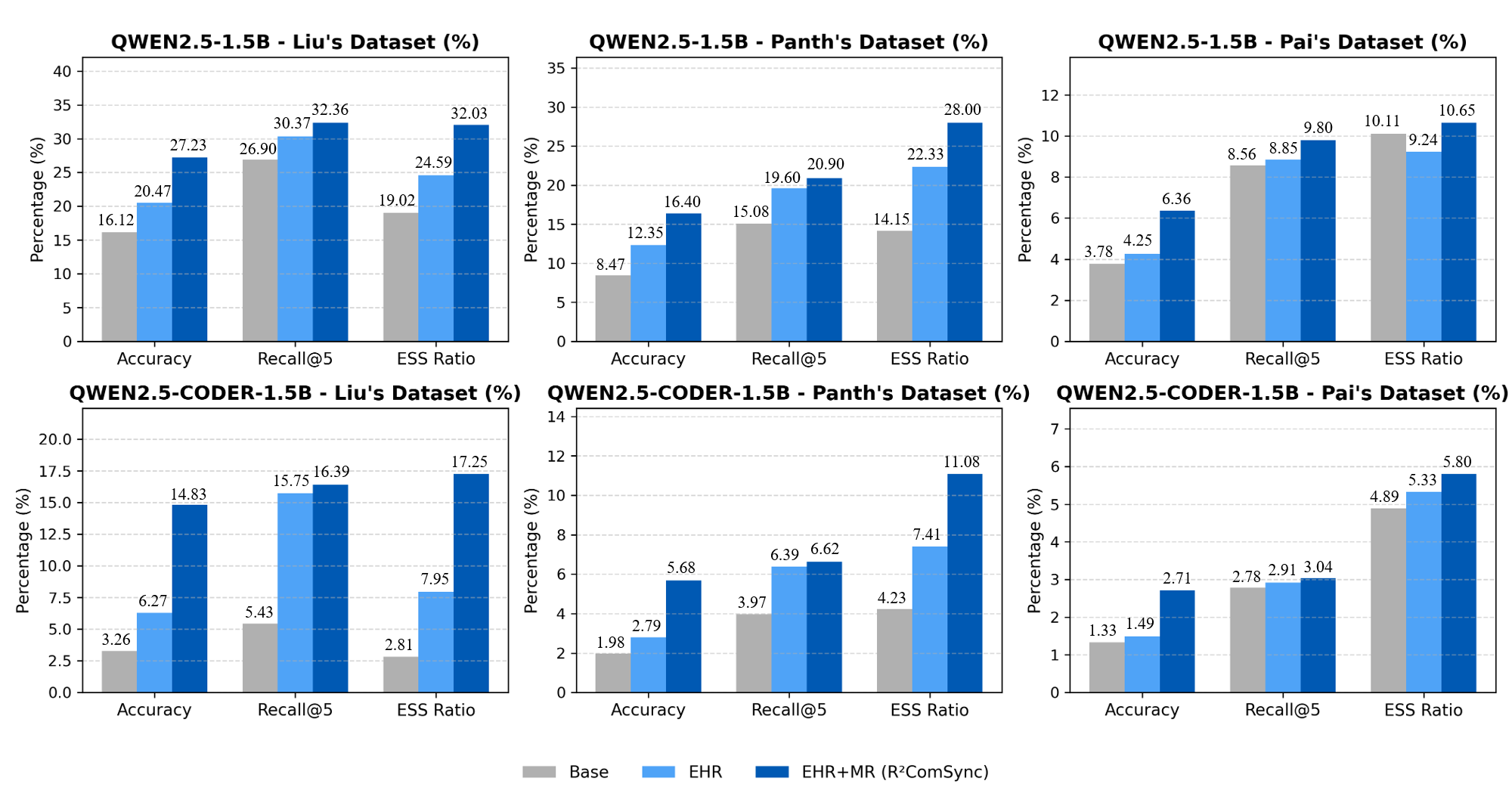}
\caption{Performance of R$^2$ComSync on Small Models.}
\label{small models}
\end{figure*}

\textbf{Objective:} In RQ2, we have investigated R$^2$ComSync's effectiveness on a series of LLMs with over 8B parameters. However, considering the resource limitations and restrictions of using these large LLMs in practice, it is imperative to examine whether R$^2$ComSync is still applicable on small LLMs.\\
\textbf{Experimental Design:} Considering that the very recently released Qwen2.5 series contains powerful LLMs with diverse parameter sizes and both code and general versions, we select two of this series (i.e., Qwen2.5-1.5B \cite{qwen25technicalreport} and Qwen2.5-Coder-1.5B \cite{hui2024qwen2}) for experiments, thereby exploring the effectiveness of R$^2$ComSync on small models. In addition, Qwen2.5-1.5B is a general LLM, while Qwen2.5-Coder-1.5B is a code LLM. Hence, we can further study performance discrepancies between code and general LLMs fairly on the CCS task.
For implementation, we download their model parameters from Huggingface to deploy them locally, and follow Section \ref{Implementaion Details} to configure hyper-parameters for LLMs. Considering the shot number ($P$) setting also affects the performance of R$^2$ComSync, we conduct multiple experiments with $P$=$\{2,4,6,8,10\}$ and average their performance for analysis.
Comparative experiments among R$^2$ComSync and its ablated versions are conducted across all three datasets, covering both Java and Python programming languages. As shown in Figure \ref{small models}, we use ``Base'' to denote small LLMs without the equipment of any of our proposals, using ``EHR'' denotes small LLMs equipped with the EHR method for ICL, and using ``EHR+MR (R$^2$ComSync)'' denotes small LLMs equipped with our whole proposal. \\
\textbf{Results:} As can be seen from Figure \ref{small models}, our proposal, i.e., R$^2$ComSync, is still effective for lifting small LLMs' performance in the CCS task. From the perspective of R$^2$ComSync as a whole, it improves Qwen2.5-1.5B by 76.77\% in terms of Accuracy, 24.27\% in terms of Recall@5, and 57.32\% in terms of ESS Ratio across all studied datasets on average. Besides, it also, on average, lifts Qwen2.5-1.5B-Coder by 213.73\%, 91.95\%, and 231.31\% in terms of each evaluation metric in order. When small LLMs are only equipped with the EHR method, their performance improvement against their Base situation is still substantial across almost all datasets and evaluation metrics. To be specific, the EHR method improves Qwen2.5-1.5B by 28.40\% in terms of Accuracy, 15.05\% in terms of Recall@5, and 26.24\% in terms of ESS Ratio across all three datasets on average. As for Qwen2.5-1.5B-Coder, the improvement becomes more significant, achieving a 48.76\% increase in Accuracy, 85.39\% in Recall@5, and 90.02\% in ESS Ratio on average across all datasets. The above analysis demonstrates the effectiveness of each of the modules we proposed on small LLMs. Horizontally comparing Qwen2.5-1.5B and Qwen2.5-1.5B-Coder, we find that the former consistently maintains a pronounced advantage. One potential explanation is that Qwen2.5-1.5B was pre-trained on a much larger corpus of 18 trillion tokens, including the code and math corpora used in Qwen2.5-1.5B-Coder \cite{qwen25technicalreport}. In contrast, the pre-training data for Qwen2.5-1.5B-Coder is only 5.2 trillion, and it uses relatively outdated data filtering techniques, which makes its capability for supervised fine-tuning less effective \cite{hui2024qwen2}. 

\begin{boxK}
 \faForward \textbf{RQ5:} R$^2$ComSync is still applicable to small models and can significantly improve their CCS capabilities as well. 
\end{boxK}

\begin{table}[htbp]
\centering
\caption{Time/GPU Memory Cost Analysis of Different Approaches on Each Dataset per Trial.}
\label{tab:time_comparison_transposed}
\renewcommand{\arraystretch}{1.15}
\setlength{\tabcolsep}{1pt}
\begin{tabular}{ccccc}
\toprule
\textbf{Model} & & \makecell{\textbf{Liu's Dataset} \\ (Train/Test/Sample)} & \makecell{\textbf{Panth's Dataset} \\ (Train/Test/Sample)} & \makecell{\textbf{Pai's Dataset} \\ (Train/Test/Sample)} \\
\midrule
CUP & \makecell[l]{Time(s)\\ Memory(GB)} &
\makecell[l]{1461.20/13.50/0.037\\ 46.5/0.642/0.642} &
\makecell[l]{ 1440.00/18.03/0.049\\  46.2/0.608/0.608} &
\makecell[l]{ 1426.90/22.10/0.074\\  47.1/0.660/0.660} \\[4pt]
\cmidrule(lr){2-5}
HEBCUP & \makecell[l]{Time(s)\\ Memory(GB)} &
\makecell[l]{—/0.21/0.0006\\ —/—/—} &
\makecell[l]{ —/0.80/0.001\\  —/—/—} &
\makecell[l]{ —/0.42/0.001\\  —/—/—} \\[4pt]
\cmidrule(lr){2-5}
CBS & \makecell[l]{Time(s)\\ Memory(GB)} &
\makecell[l]{ 1619.70/8.08/0.022\\ 46.5/0.642/0.642} &
\makecell[l]{ 1459.00/15.19/0.021\\  46.2/0.608/0.608} &
\makecell[l]{ 1444.00/18.06/0.060\\ 47.1/0.660/0.660} \\[4pt]
\midrule
\multicolumn{5}{c}{\cellcolor{gray!20}\textbf{R²ComSync Scenarios}}\\\midrule
 Llama3-70B & \makecell[l]{Time(s)\\ Memory(GB)} &
 \makecell[l]{ —/2313.27/6.286 \\ —/145.352/145.352} &
\makecell[l]{ —/4582.74/6.227\\  —/144.847/144.847} &
\makecell[l]{ —/3704.97/12.350\\ —/149.938/149.938} \\[4pt]
\cmidrule(lr){2-5}
 Llama3-8B  & \makecell[l]{Time(s)\\ Memory(GB)} &
 \makecell[l]{ —/334.85/0.910 \\ —/17.075/17.075} &
\makecell[l]{ —/696.28/0.946\\  —/16.958/16.958} &
\makecell[l]{ —/670.60/2.235\\ —/18.023/18.023} \\[4pt]
\cmidrule(lr){2-5}
 GPT-3.5-Turbo    & \makecell[l]{Time(s)\\ Memory(GB)} &
\makecell[l]{ —/453.60/1.233 \\ —/—/—} &
\makecell[l]{ —/882.83/1.199\\ —/—/— } &
\makecell[l]{ —/523.20/1.744\\ —/—/—} \\[4pt]
\cmidrule(lr){2-5}
 Qwen2.5-1.5B     & \makecell[l]{Time(s)\\ Memory(GB)} &
\makecell[l]{ —/130.77/0.355 \\ —/9.271/9.271} &
\makecell[l]{ —/223.13/0.303\\—/8.643/8.643} &
\makecell[l]{ —/230.81/0.769\\ —/14.915/14.915} \\[4pt]
\cmidrule(lr){2-5}
 Qwen2.5-Coder-1.5B & \makecell[l]{Time(s)\\ Memory(GB)} &
 \makecell[l]{ —/159.59/0.434 \\ —/9.271/9.271} &
\makecell[l]{ —/299.19/0.407\\—/8.643/8.643} &
\makecell[l]{ —/255.09/0.850\\ —/14.915/14.915} \\[4pt]
\bottomrule
\end{tabular}
\smallskip
\parbox{\textwidth}{\footnotesize \textit{Note:}  The reported GPU memory usage represents the theoretical value because we employ vLLM for inference acceleration, using a fixed memory allocation of 48×4×0.95 GB for LLMs.}
\end{table}

\begin{table}[htbp]
\centering
\caption{Total Token Consumption and Cost of GPT-3.5-Turbo on Each Dataset per Trial.}
\label{tab:token_cost}
\renewcommand{\arraystretch}{1.2}
\setlength{\tabcolsep}{11pt}
\begin{tabular}{lcccc}
\toprule
\makecell{\textbf{Liu's Dataset} \\ (Input/Output/Total Cost)} &\makecell{\textbf{Panth's Dataset}\\ (Input/Output/Total Cost)} & \makecell{\textbf{Pai's Dataset}\\ (Input/Output/Total Cost)} \\
\midrule
195158.2/12641.0/\$0.052 & 360,315.4/29,172.6/\$0.060 & 278,088.6/52,411.4/\$0.060 \\

\bottomrule
\end{tabular}
\smallskip
\parbox{\textwidth}{\footnotesize \textit{Note:} The cost of GPT-3.5-Turbo is calculated based on \$0.25 per million input tokens and \$0.75 per million output tokens\cite{islam2025evaluating}.}
\end{table}

\section{Discussion}
\label{Discussion}

\subsection{Cost Analysis}
Table \ref{tab:time_comparison_transposed} showcases the GPU memory and time cost of each studied approach for each dataset per trial, where they are further partitioned into ``Train'', ``Test'', and ``Sample''.  ``Train'' and ``Test'' denote the total cost during the training and testing stage, respectively. ``Sample'' denotes the cost of running each sample during the testing stage, thereby providing a reference for professionals for their practical usage. It should be noted that the testing stage follows a sample-by-sample inference; thus, the total GPU memory consumption in ``Test'' and ``Sample'' is the same. Besides, we also exclude HatCUP and Toper, due to their non-reproducibility, as mentioned in Section \ref{Motivation}.
As can be seen, HEBCUP, as a heuristic-based approach without GPU support, consumes minimal computational resources per trial. The time and memory costs of R$^2$ComSync include retrieval, inference, and re-ranking phases. Since LLMs in R$^2$ComSync are mandated for direct inference, their training cost is 0. Besides, in our experiments, shot number ($P$) has been tested from 2 to 10 for R$^2$ComSync, we average the time and memory cost across diverse shot numbers, owing to their significant influence on cost. Apparently, since LLMs normally carry massive parameters, their memory and time costs are higher than existing approaches overall. However, their cost is still acceptable, especially for a single sample when making inferences. Except for Llama3-70B, R$^2$ComSync with most LLMs can complete a synchronization within 1 second, while still keeping substantial superiority against existing approaches overall, as shown in Section \ref{Effectiveness of R$^2$ComSync (RQ2)} and \ref{Applicability of R$^2$ComSync to small models (RQ5)}.  
As for GPT-3.5-Turbo, since it was invoked from OpenAI's APIs rather than deployed locally, we cannot obtain its GPU memory consumption. However, we introduce another cost analysis perspective: token consumption, shown in Table \ref{tab:token_cost}, where both the input and output token consumption are recorded in detail for reference.

\subsection{Case Study}
\label{Case Study}
\begin{figure}[htbp]
\centering
\includegraphics[width=0.6\textwidth]
{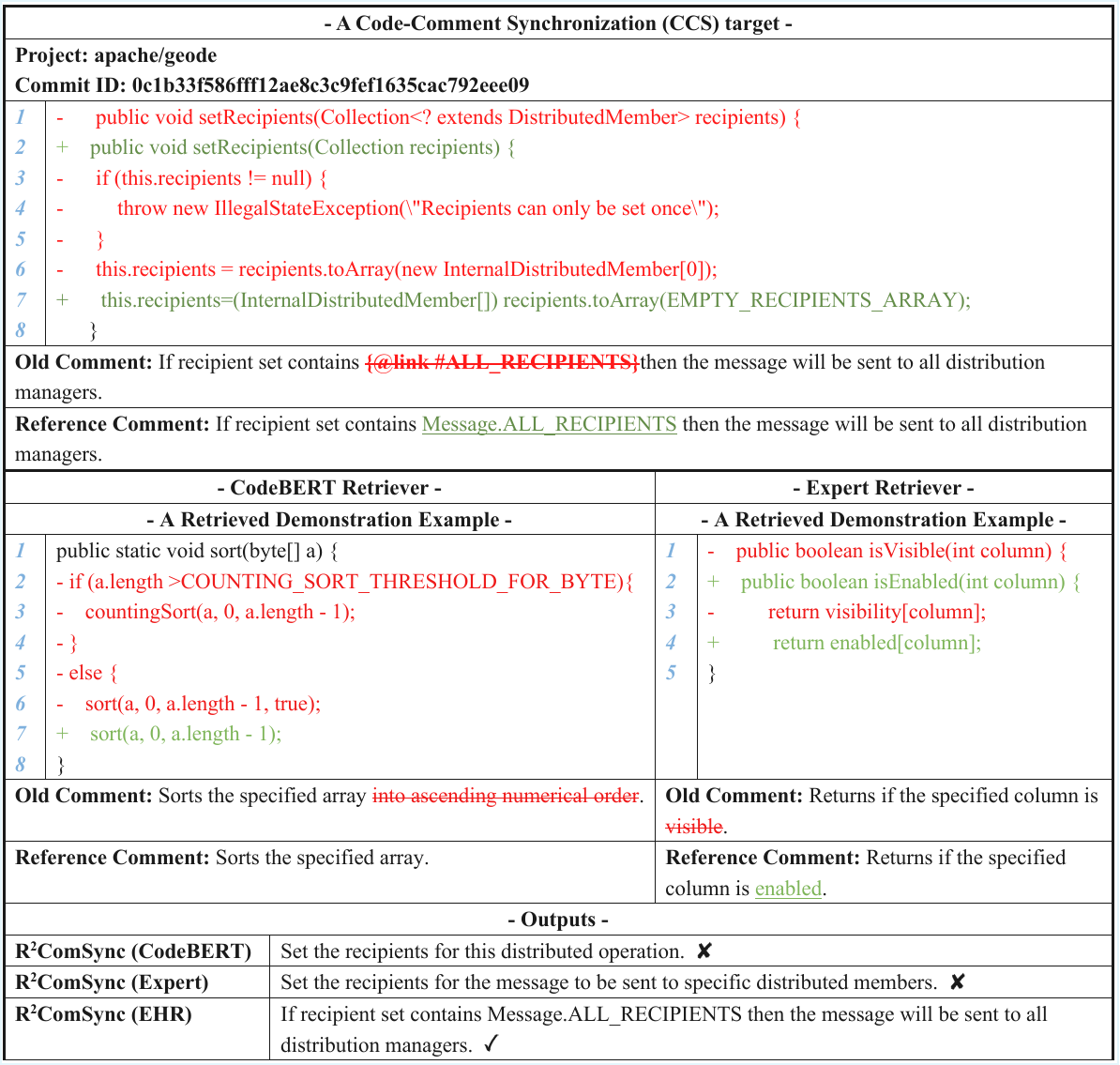}
\caption{Case Study 1, Illustrating the Strength of the EHR Method.}
\label{Case Study 1}
\end{figure}

\begin{figure}[htbp]
\centering
\includegraphics[width=0.6\textwidth]
{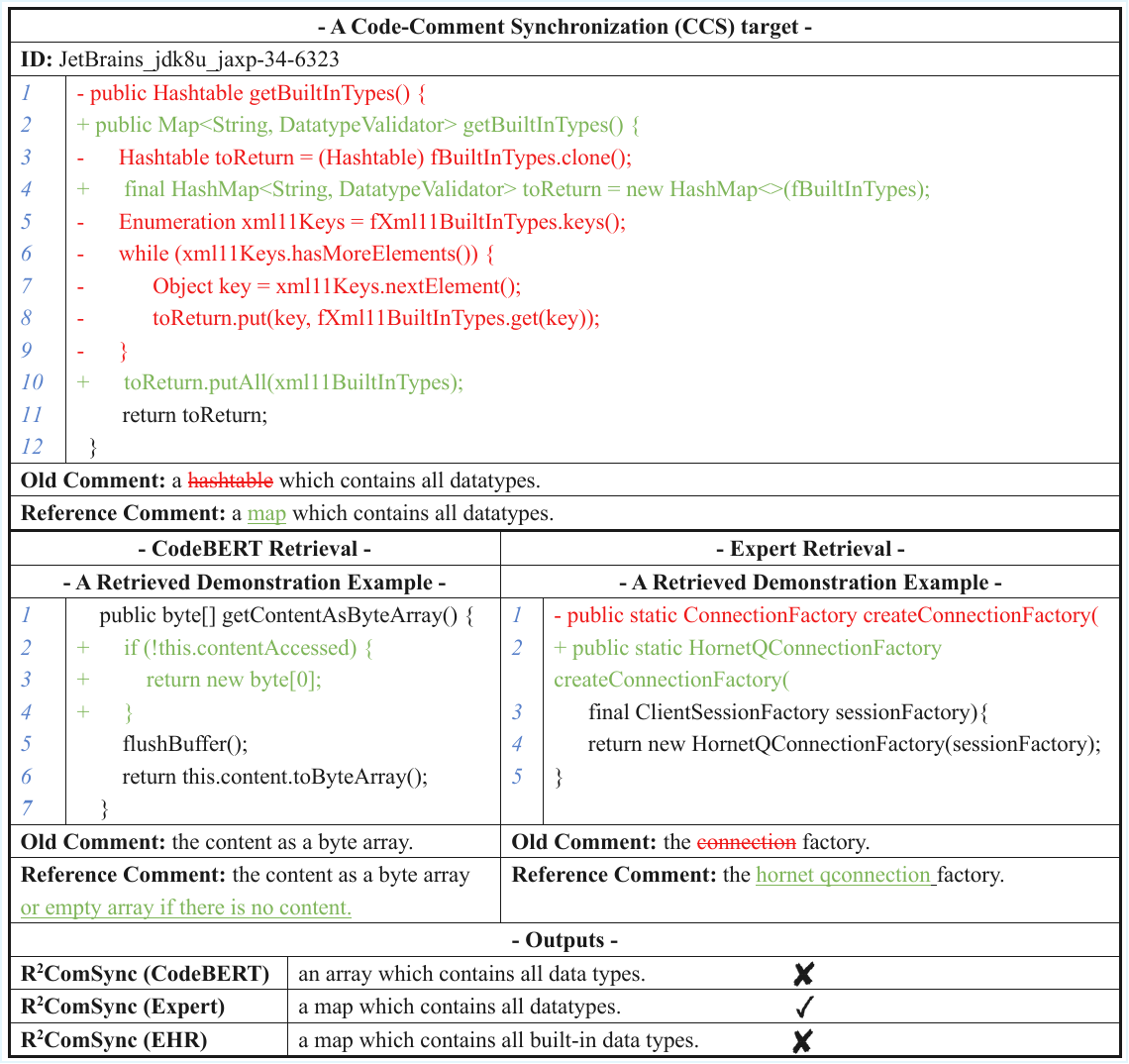}
\caption{Case Study 2, Illustrating the Limitation of the EHR Method.}
\label{Case Study 2}
\end{figure}

\begin{figure}[htbp]
\centering
\includegraphics[width=0.6\textwidth]
{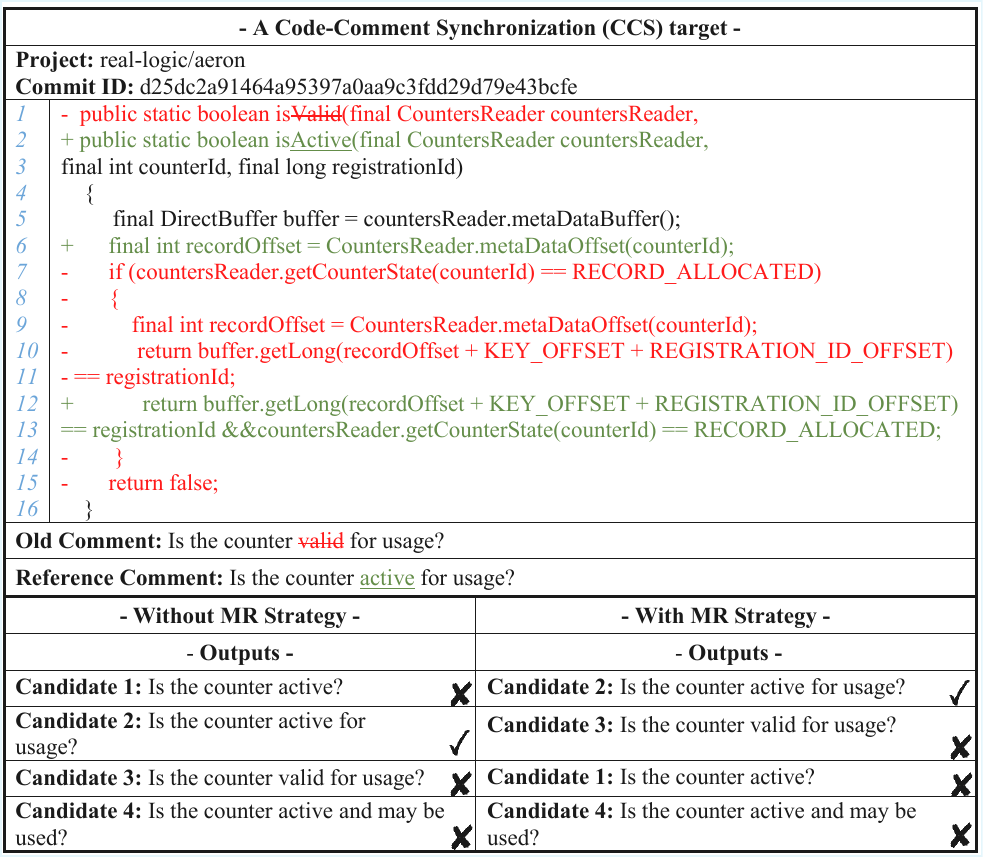}
\caption{Case Study 3, Illustrating the Strength of the MR Strategy.}
\label{Case Study 3}
\end{figure}

Figures \ref{Case Study 1}, \ref{Case Study 2}, and \ref{Case Study 3} present three typical case studies that demonstrate the effectiveness of each proposed module. For clarification, we use red (green) lines in the code snippet to denote deleted (newly added) statements, and we also highlight the removed (newly added) tokens in red (green) for old (new) comments. 

In Case Study 1 (Figure \ref{Case Study 1}), we showcase an example from apache/geode. In the original code, the condition ``if (this.recipients != null)'' is used to ensure that ``recipients'' can only be set once, throwing an ``IllegalStateException'' if violated. In the updated new code, however, this check has been removed, allowing ``recipients'' to be set repeatedly. Additionally, the method for array conversion has been modified: the original code uses ``recipients.toArray(new InternalDistributedMember[0])'', whereas the new code performs an explicit cast with ``(InternalDistributedMember[]) recipients.toArray(EMPTY\_RECIPIENTS\_ARRAY)''. These two changes significantly affect the semantics of the code.
Given this CCS target, we use CodeBERT and expert features to retrieve the most similar example, respectively. As can be seen, CodeBERT-retrieved example clearly eliminates the threshold-checking logic ``if (a.length\(>\) COUNTING\_SORT\_THRESHOLD\_FOR\_BYTE)'' which is semantically the same as the cod changes in the CCS target. As for the expert-retrieved example, it carries similar code-comment change patterns in terms of the change complexity and involvement. However, neither of them can make LLMs generate \textit{correct synchronization} solely. But when we gather these two examples together for ICL, we obtain the final correct result. 



In Case Study 2 (Figure \ref{Case Study 2}), the CCS target modifies the return type from ``Hashtable'' to the generic ``Map\(<\)String, DatatypeValidator\(>\)'' and accordingly makes some follow-up changes in code, such as optimizing collection operations and constituting a type-safety enhancement. The above changes are precisely reflected in comments, i.e., change ``hashtable'' to ``map''. 
CodeBERT retrieved an example featuring logical augmentation through adding conditional checks, encompassing array operations. Semantically, the CCS target is similar to this example, as ``Array'' intuitively contains a relatively similar representation to ``Map'' and ``Hashtable'', owing to their data structure usage. However, this example cannot instruct LLMs to their correct results in new comments (see the third-last row), because of the absolutely different code-comment change patterns. 
In contrast, the expert-retrieved example follows a similar change pattern to the CCS target: the modification is induced by the change of return type, i.e., from the ``ConnectionFactory'' to ``HornetQConnectionFactory'', which is also reflected in its old and new comments. As such, expert retrieval offers a more instructive demonstration and makes LLMs generate the correct results (see the second-last row). 

Case Study 3 (Figure \ref{Case Study 3}) is a CCS sample in Liu's dataset. In this case, LLMs generate four candidate results for the given CCS target. Yet, the \textit{correct synchronization} is ranked second without the MR strategy, as shown in the lower-left corner of Figure \ref{Case Study 3}. Subsequently, we conduct the MR strategy on the generated candidates and list the re-ranking results in the lower-right corner of Figure \ref{Case Study 3}. Specifically, Candidates 1 \footnote{Rule 3 ($\epsilon$) of Candidate 1: $ED_{cm}/N_{cm_{s}}$(3/7)=0.429$>\epsilon$=0.25} and 4 \footnote{Rule 3 ($\epsilon$) of Candidate 4: $ED_{cm}/N_{cm_{s}}$(5/7)=0.714$>\epsilon$=0.25} both violate Rule 3 because their edit distances relative to the old comment exceed our predefined threshold. Since it is a severe mistake, they are ranked third and fourth. Towards Candidate 3, the token ``valid'' is not changed by the token ``active'', which means it does not sense the code change occurring in the function name. That is to say, Candidate 3 violates Rule 1. As Rule 1 is a weak mistake, it consequently ranks second. To this end, Candidate 2 is finally prioritized among other candidates and ranks first, as it does not violate any rule. In this way, the MR strategy takes effect in the re-ranking phase of our proposal. 

\begin{figure}[htbp]
\centering
\includegraphics[width=0.6\textwidth]
{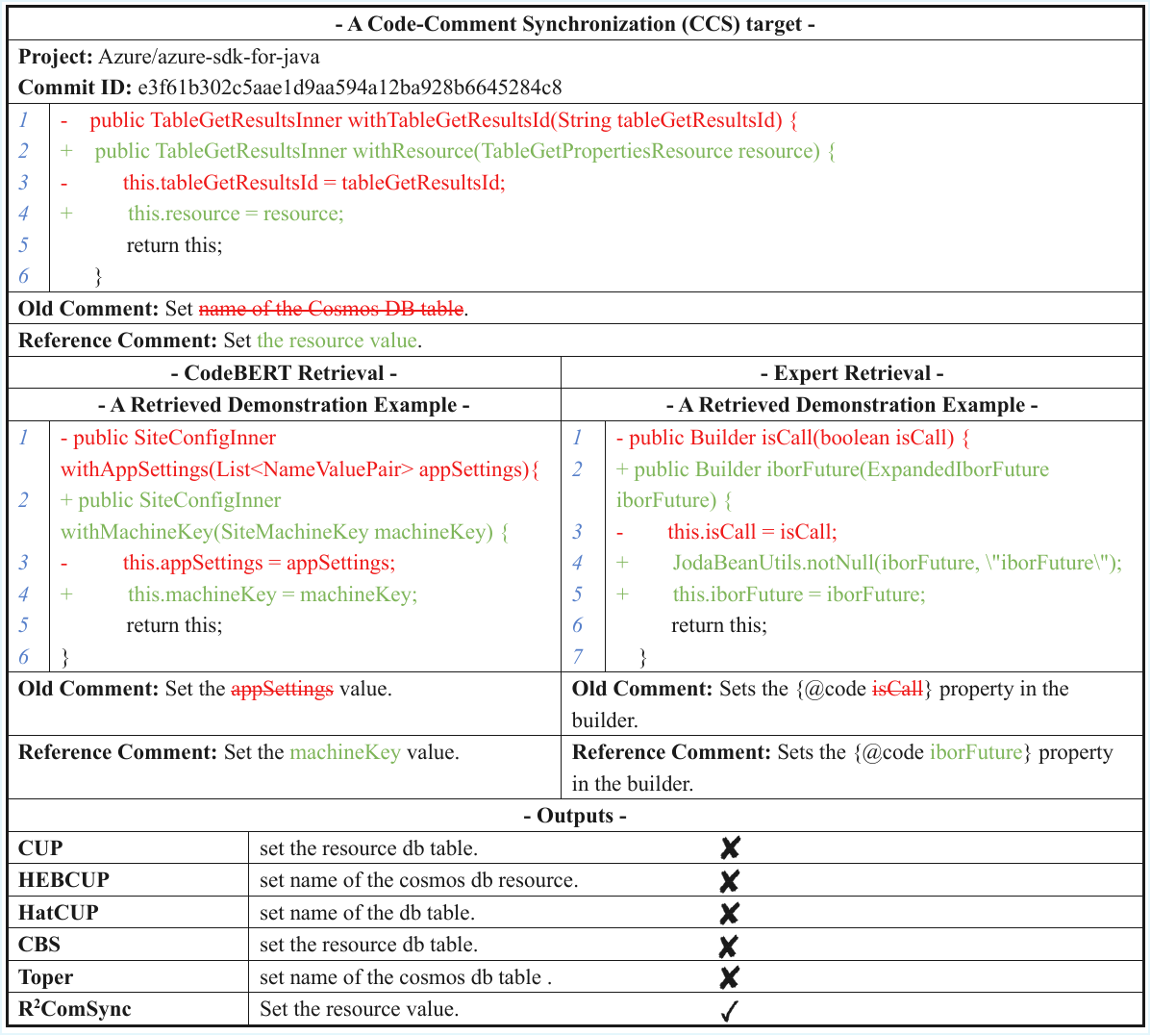}
\caption{Case Study 4: Comparison between R$^2$ComSync and Baseline.}
\label{Case Study 4}
\end{figure}

Owing to the advantages of our proposed modules in R$^2$ComSync as illustrated above, R$^2$ComSync makes LLMs becomes competitive and even substantially exeling existing approaches. Case 4 is a typical example, shown in Figure \ref{Case Study 4}. The code changes in this CCS target  modifies the setting value from ``tableGetResultsId'' to ``resource'', which is also precisely reflected in the comment changes. As can be seen, none of the existing approach achieves the \textit{correct synchronization}. For example, heuristic-based HEBCUP only handled the replacement between ``table'' and ``resource''. Deep learning-based CUP struggles in how to deal with the relationship between ``resource'' and ``db table''. R$^2$ComSync retrieves similar demonstrations from code-comment semantics and change patterns for ICL, leading to its successful synchronization for this case.  

\subsection{Where does R$^2$ComSync Fail}
\label{Where does R$^2$ComSync Fail}

\begin{figure}[htbp]
\centering
\includegraphics[width=0.6\textwidth]
{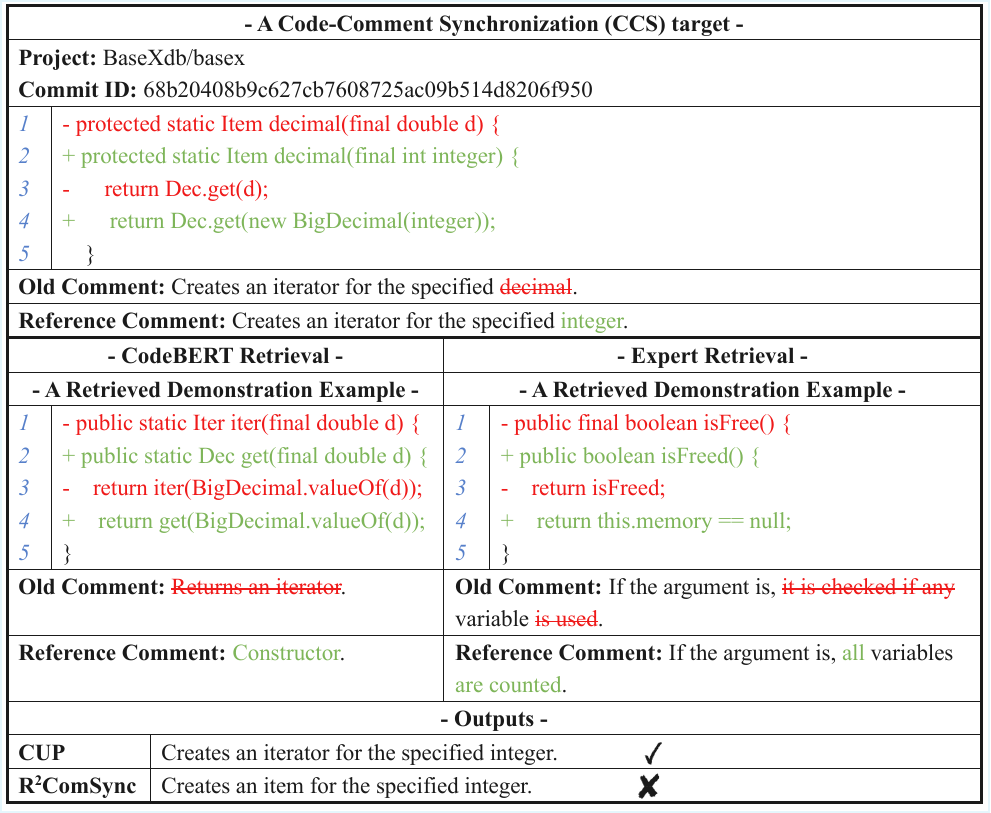}
\caption{Case Study 5, A Sample which R$^2$ComSync Fail.}
\label{Case Study 5}
\end{figure}


Although R$^2$ComSync exhibits powerful CCS ability, it has limitations and failures in some cases. This section carefully inspects a series of failed examples that R$^2$ComSync cannot generate \textit{correct synchronization}s and summarizes three main situations:

\textbf{(1) Misleading demonstrations.} 
Case Study 5 (Figure \ref{Case Study 5}) presents an example where the training-based CUP succeeds while our method fails. R$^2$ComSync produces a semantically incorrect new comment, i.e., the term ``item'' does not match the intended ``iterator''.
However, the error does not stem from the key code changes involving precision optimization from ``double'' to ``BigDecimal'', but rather from the substitution of ``iterator'' with ``item''. However, based on the retrieved demonstrations from the CCS target and CodeBERT or expert aspects, ``item'' has never appeared in the input context of R$^2$ComSync. Hence, we speculate it is a hallucination based on LLMs' own internal distribution. The reason for this hallucination is potentially due to misleading demonstrations. As can be seen, although CodeBERT retrieval obtains a semantically similar example, its code-comment change patterns differ from those of the CCS target. As for the expert retrieval, it also does not offer proper examples for ICL. In other words, our prepared demonstration pools have no instructive examples from the two perspectives mentioned above. As such, R$^2$ComSync fails on this case.

In contrast, CUP, being directly fine-tuned on in-domain data, exhibits stronger conservatism toward domain-specific terminology. Although it is less flexible, its internal distribution aligns more closely with the target domain, preventing arbitrary hallucinations. It is conceivable that incorporating CUP-generated examples into the retrieval corpus could help LLMs focus more consistently on the same domain and thereby reduce hallucinations. While this would increase computational cost, it represents a promising direction for future research.

Besides, developers' intentions when updating comments sometimes are elusive. Especially, when a great chunk of code is changed, various updated information has the probability of being reflected in the new comments, making it hard to precisely capture the developers' updating intention by representations of code changes and old comments. To this end, the CCS demonstrations retrieved by the above representations may mislead the targets, such as case \cite{Improved52:online}.

\textbf{(2) Specious candidates.} 
A specious candidate does not violate any rule we predefined, but is not identical to the reference comment. Typically, these candidates normally correctly update partial comments or make relatively tiny mistakes when synchronization. In this case, once its initial ranking is high, the heuristic-based MR strategy cannot detect it and re-rank its order. For example, when handling cases \cite{Settings80:online, Fixedgec54:online}, R$^2$ComSync misses a Javadoc markup ``@code'' but correctly updates all other contents, leaving specious candidates.

\textbf{(3) Misjudgments of automatic evaluation.}  
Automatic evaluation cannot identify synonyms or paraphrases when assessing the qualities of generated new comments. Thus, some cases are considered correct via human evaluation but are deemed flawed in automatic evaluation, such as case \cite{Renamequ2:online}, where the generated new comment uses ``is not'' instead of ``no'' in its reference, causing the misjudgment. 

\subsection{Implications}
This section introduces the implications of our work from the perspectives of researchers and practitioners, respectively.

\textbf{Implications for researchers.}
This work, for the first time, revealed the weakness of LLMs in handling the CCS task against existing SOTA approaches. Compared with recent work on the CCS task, such as \cite{fan2024exploring}, our work introduces the motivation of further enhancing LLMs instead of only empirically discussing their capabilities. Besides, as the first attempt to unleash LLMs' ability in CCS, we design an EHR method to retrieve effective ICL demonstrations and propose novel re-ranking rules to prioritize correct-prone candidates. Although ICL and re-ranking are common methods in LLM-augmented approaches, we provide novel thoughts for the above methods in the domain of CCS research. For example, we combine code-comment semantics and change patterns for relevant demonstration retrieval, where the former can provide synchronization directions in a macro aspect, while the latter can offer detailed synchronization guidance on specific patterns (See Section \ref{Contributions of Ensemble Hybrid Retrieval (RQ3)} for details), making our retrieval method different from prior retrieval-augmented LLM approaches. On the other hand, we also summarize a series of re-ranking rules via large-scale analysis on CCS samples, further improving the CCS performance of R$^2$ComSync as shown in the Section \ref{Contributions of Multi-turn Re-ranking Strategy (RQ4)}. As such, the academic novelty of R$^2$ComSync lies more in its specific and targeted design for the CCS task, rather than in applying other methods directly to this area. 
In addition, we also discuss a series of typical failures that R$^2$ComSync exhibited in our experiments (see Section \ref{Where does R$^2$ComSync Fail} for details), thereby providing insights for future improvement and research directions.

\textbf{Implications for practitioners.}
This paper proposed R$^2$ComSync, an effective LLM-augmented tool to synchronize comments to their corresponding code snippets. To help practitioners understand its use in daily development and maintenance, we thoroughly investigated R$^2$ComSync on both real-world Java and Python projects to assess its effectiveness in Section \ref{Effectiveness of R$^2$ComSync (RQ2)}. Besides, we also provided a series of actionable insights on the configuration of the EHR method and the MR strategy of R$^2$ComSync (See Section \ref{Contributions of Ensemble Hybrid Retrieval (RQ3)} and \ref{Contributions of Multi-turn Re-ranking Strategy (RQ4)}). For example, we discussed the weight allocation of CodeBERT retrieval and expert retrieval in the EHR method for different data samples, empirically highlighting the superiority of change-pattern-aligned demonstrations in short CCS samples and their low dependency on data distribution consistency. Besides, we also examined appropriate settings for shot numbers when retrieving ICL demonstrations. In addition, the utility of each re-ranking rule and its hyper-parameter settings was also studied to offer practical suggestions. Moreover, in Section \ref{Discussion}, we further explored the cost of R$^2$ComSync in practical usage and showcased both its successful and failed cases, intuitively illustrating its advantages and precautions for practitioners.

\subsection{Threats to Validity}
There are mainly four threats to the validity of this work.

\textbf{Potential data leakage.}
Data leakage occurs when testing samples are visible during model training, leading to an overestimation of model performance in experiments, which is also a type of data snooping problem. Our proposal is experimented on Llama3, GPT3.5, and Qwen series, which are LLMs trained on open-source code repositories that are not publicly disclosed. As such, there is a potential threat that they may have seen both before- and after-change code in our testing set during their pre-training, causing their memorization in certain CCS samples. Nonetheless, all studied LLMs were pre-trained on GitHub code files using Causal Language Modeling (CLM), which is entirely different from the inference pattern of the CCS task. This task involves using old and new code, along with old comments, as inputs, while designating new comments as outputs. This discrepancy can be categorized as an unpaired contamination, which has limited effects on the follow-up inference on the CCS task as revealed in \cite{yang2025rethinking}.   
Besides, No LLM performs well in the zero-shot setting compared to SOTA baselines (refer to Table \ref{Performances of Baselines and LLMs.}), indicating that these LLMs have not memorized the testing samples. Instead, our proposed R$^2$ComSync significantly improves the performance of LLMs, demonstrating the effectiveness of our proposal. To this end, we believe this threat is minimal.

\textbf{The generalizability of our experimental results.}
In this work, we conducted experiments on three real-world CCS datasets across two programming languages, including the statically-typed Java and dynamically-typed Python, explicitly demonstrating the generality of R$^2$ComSync. Although our studied datasets are mainly on the function level, it is the most prevalent comment updating granularity in practical development \cite{liu2020automating, lin2021automated, zhu2022HatCUP, yang2021significance, lin2022predictive,liu2021just}.
Besides, we exhaustively select all state-of-the-art approaches in this field as our baselines for comparison and conduct both automatic and human evaluations for comprehensive assessments. Therefore, the threat of generalizability is limited.

\textbf{CCS Trials on other LLMs.}
In this paper, we conduct CCS experiments with Llama3-8/70B, GPT3.5, Qwen2.5, and Qwen2.5-Coder, yet there are other alternatives we have not experimented on, such as CodeGen \cite{nijkamp2022codegen}, InCoder \cite{fried2022incoder}, and DeepSeek series \cite{liu2024deepseek,guo2024deepseek}. Nevertheless, our selected LLMs contain both closed-source (e.g., GPT-3.5-Turbo) and open-source LLMs (e.g., Llama3); code (e.g., Qwen2.5-Coder-1.5B) and general (e.g., Qwen2.5-1.5B) LLMs; as well as large (e.g., Llama3-70B) and small LLMs (e.g., Qwen2.5-1.5B) of different families. Therefore, we believe the choice of studied LLMs is adequate and comprehensive to a great extent.

\textbf{Limited experiments on hyper-parameter settings.}
The hyper-parameter used in R$^2$ComSync involves shot number ($P$) and two thresholds of Rule 2($\sigma$) and 3($\epsilon$). Although we have tested them within effective ranges based on our usage experience on R$^2$ComSync and achieved SOTA performance against existing baselines, other hyper-parameter combinations might yield better performance. However, owing to the restriction of computational resources, we leave a broader exploration of hyper-parameters for future work.


\section{Conclusion}
This paper introduces the R$^2$ComSync approach, aiming to automate the synchronization of code and comments to reduce the workload of developers during software maintenance and evolution. Specifically, we propose a hybrid ensemble retrieval method to collect similar CCS examples, focusing on both the semantics of code and comments as well as the patterns of changes. This method provides precise guidance for the generation objectives of LLMs. To enhance the effectiveness of our approach, we also propose a multi-turn re-ranking strategy to prioritize correct-prone candidates generated by LLMs. Extensive experiments have demonstrated that R$^2$ComSync significantly lifts LLMs' capability in the CCS task and achieves the SOTA performance among diverse existing approaches. In the future, we will further explore more effective methodologies to advance in this area. For example, designating LLMs to summarize code changes as intermediate guidance for synchronizing old comments to their new versions. 

\section{Declarations section}
\subsection{Funding}
This work was partially supported by the National Natural Science Foundation of China under Grant U24B20149-2 and in part by the Natural Science Foundation of Shandong Province under Grant ZR2024QF093.
\subsection{Ethical approval}
Not Applicable.
\subsection{Informed Consent}
All participants who joined the human evaluation were provided with an informed consent form outlining the study’s objectives, procedures, potential risks, and benefits. Participation was voluntary, and respondents could withdraw at any time without consequences. 
\subsection{Author Contributions}
\textbf{Zhen Yang:} Conceptualization, Methodology, Software, Formal analysis, Validation, Writing. \textbf{Hongyi Lin:} Software, Formal analysis, Investigation, Data curation, Writing. \textbf{Xiao Yu:} Conceptualization, Validation, Writing. \textbf{Jacky Wai Keung:} Supervision, Conceptualization, Validation. \textbf{Shuo Liu:} Software, Formal analysis, Investigation, Data curation, Writing. \textbf{Pak Yuen Patrick Chan:} Validation, Writing. \textbf{Yicheng Sun:} Validation, Writing. \textbf{Fengji Zhang:} Validation, Writing.
\subsection{Data Availability Statement}
\subsection{Conflict of Interest}
The authors declared that they have no conflict of interest.
\subsection{Clinical Trial Number in the manuscript}
Not Applicable.




\bibliography{sn-bibliography}

\end{document}